\documentclass{aa}

\usepackage{graphicx}
\usepackage{txfonts}
\usepackage{natbib}     %A&A reference needed usepackage
\usepackage{xcolor} %for color text
\usepackage{threeparttable} %for table note

\bibpunct{(}{)}{;}{a}{}{,} % for citation style,to follow the A&A style

\begin{document}

   \title{Ionized calcium in the atmospheres of two ultra-hot exoplanets WASP-33b and KELT-9b}

   \author{F. Yan\inst{1}
   		  \and
   		  N. Casasayas-Barris\inst{2,3}
   		  \and
          K. Molaverdikhani\inst{4}
          \and
          F.~J. Alonso-Floriano\inst{5}
          \and
          A. Reiners\inst{1}
          \and
          E. Pall\'e\inst{2,3}
          \and
          Th. Henning\inst{4}
          \and
          P.~Molli\`ere\inst{5}
          \and
          G.~Chen\inst{6}
          \and
          L.~Nortmann\inst{2,3}
          \and
          I.~A.~G.~Snellen\inst{5}
          \and
          I.~Ribas\inst{7,8},  A.~Quirrenbach\inst{9}, J.~A.~Caballero\inst{10}, P.~J.~Amado\inst{11},    
          M.~Azzaro\inst{12}, F.~F.~Bauer\inst{11}, M.~Cort\'es Contreras\inst{10}, S. Czesla\inst{13}, S.~Khalafinejad\inst{9}, L.~M.~Lara\inst{11}, M.~L\'opez-Puertas\inst{11}, 
		D. Montes\inst{14}, E.~Nagel\inst{13}, M.~Oshagh\inst{1}, A.~S\'anchez-L\'opez\inst{11}, M.~Stangret\inst{2,3}
		 \and
		 M. Zechmeister\inst{1}
          }
          
   	\institute{Institut f\"ur Astrophysik, Georg-August-Universit\"at, Friedrich-Hund-Platz 1, 37077 G\"ottingen, Germany\\
	\email{fei.yan@uni-goettingen.de}
	\and
	Instituto de Astrof{\'i}sica de Canarias (IAC), Calle V{\'i}a Lactea s/n, 38200 La Laguna, Tenerife, Spain
\and
Departamento de Astrof{\'i}sica, Universidad de La Laguna, 38026  La Laguna, Tenerife, Spain
\and
Max-Planck-Institut f{\"u}r Astronomie, K{\"o}nigstuhl 17, 69117 Heidelberg, Germany
\and
	Leiden Observatory, Leiden University, Postbus 9513, 2300 RA, Leiden, The Netherlands
\and
Key Laboratory of Planetary Sciences, Purple Mountain Observatory, Chinese Academy of Sciences, Nanjing 210033, China
\and
Institut de Ci\`encies de l'Espai (CSIC-IEEC), Campus UAB, c/ de Can Magrans s/n, 08193 Bellaterra, Barcelona, Spain
\and
Institut d'Estudis Espacials de Catalunya (IEEC), 08034 Barcelona, Spain
\and
Landessternwarte, Zentrum f\"ur Astronomie der Universit\"at Heidelberg, K\"onigstuhl 12, 69117 Heidelberg, Germany
\and
Centro de Astrobiolog{\'i}a (CSIC-INTA), ESAC, Camino bajo del castillo s/n, 28692 Villanueva de la Ca{\~n}ada, Madrid, Spain
\and
Instituto de Astrof{\'i}sica de Andaluc{\'i}a (IAA-CSIC), Glorieta de la Astronom{\'i}a s/n, 18008 Granada, Spain
\and
Centro Astron{\'o}nomico Hispano-Alem{\'a}n (CSIC-MPG), Observatorio Astron{\'o}nomico de Calar Alto, Sierra de los Filabres, E-04550 G{\'e}rgal, Almer\'ia, Spain
\and
Hamburger Sternwarte, Universit{\"a}t Hamburg, Gojenbergsweg 112, 21029 Hamburg, Germany
\and
%Th{\"u}ringer Landessternwarte Tautenburg, Sternwarte 5, 07778 Tautenburg, Germany
%\and
Departamento de F\'{i}sica de la Tierra y Astrof\'{i}sica 
and IPARCOS-UCM (Intituto de F\'{i}sica de Part\'{i}culas y del Cosmos de la UCM), 
Facultad de Ciencias F\'{i}sicas, Universidad Complutense de Madrid, E-28040, Madrid, Spain
%\and
%Centro de Astrobiolog{\'i}a, Carretera de Ajalvir km 4, E-28850 Torrej{\'o}n de Ardoz, Madrid, Spain
	}
        \date{Received 29 July 2019; Accepted 30 October 2019}
  % \date{Received September 15, 1996; accepted March 16, 1997}

% \abstract{}{}{}{}{}
% 5 {} token are mandatory

  \abstract
 {Ultra-hot Jupiters are emerging as a new class of exoplanets. Studying their chemical compositions and temperature structures will improve the understanding of their mass loss rate as well as their formation and evolution.
We present the detection of ionized calcium in the two hottest giant exoplanets -- KELT-9b and WASP-33b. 
By utilizing transit datasets from CARMENES and HARPS-N observations, we achieved high confidence level detections of \ion{Ca}{ii} using the cross-correlation method.
We further obtain the transmission spectra around the individual lines of the \ion{Ca}{ii} H\&K doublet and the near-infrared triplet, and measure their line profiles. The \ion{Ca}{ii} H\&K lines have an average line depth of 2.02 $\pm$ 0.17 $\%$ (effective radius of 1.56 $R_\mathrm{p}$) for WASP-33b and an average line depth of 0.78 $\pm$ 0.04 $\%$ (effective radius of 1.47 $R_\mathrm{p}$) for KELT-9b, which indicates that the absorptions are from very high upper atmosphere layers close to the planetary Roche lobes. The observed \ion{Ca}{ii} lines are significantly deeper than the predicted values from the hydrostatic models. Such a discrepancy is probably a result of hydrodynamic outflow that transports a significant amount of \ion{Ca}{ii} into the upper atmosphere.
The prominent \ion{Ca}{ii} detection with the lack of significant \ion{Ca}{i} detection implies that calcium is mostly ionized in the upper atmospheres of the two planets.
 }

   \maketitle

%
%________________________________________________________________

\section{Introduction}
Ultra-hot Jupiters (UHJs) are a new class of exoplanets emerging in the recent years. They are highly irradiated gas giants with day-side temperatures that are typically  $\gtrsim$ 2200 K \citep{Parmentier2018}. Most of these planets orbit very close to A- or F-type stars.
Their extremely high day-side temperatures cause thermal dissociation of molecules and ionization of atoms \citep{Arcangeli2018, Lothringer2018}. Depending on the heat transport efficiency, different chemical components can form at their night-sides as well as terminators \citep{Parmentier2018, Bell2018, Helling2019}.
Furthermore, the strong stellar ultraviolet (UV) and/or extreme-ultraviolet irradiation causes significant mass loss, affecting the planetary atmospheric composition and evolution \citep{Bisikalo2013, Fossati2018}.

Observations of UHJs have revealed peculiar properties of their atmospheres. For example, \cite{Kreidberg2018} found the absence of $\mathrm{H_2O}$ features at the day-side atmosphere of WASP-103b and they attributed this to the thermal dissociation of $\mathrm{H_2O}$. \cite{Arcangeli2018} analyzed the day-side spectrum of WASP-18b and found that molecules are thermally dissociated while the $\mathrm{H^-}$ ion opacity becomes important.
\cite{Yan2018} detected strong hydrogen $\mathrm{H\alpha}$ absorption in the transmission spectrum of KELT-9b, which indicates that the planet has a hot escaping hydrogen atmosphere. The $\mathrm{H\alpha}$ line was also detected in two other UHJs: MASCARA-2b \citep{Casasayas-Barris2018} and WASP-12b \citep{Jensen2018}.
Some atomic/ionic metal lines are also detected in UHJs, for instance, \cite{Fossati2010} detected \ion{Mg}{ii} in WASP-12b using UV transmission spectroscopy with the \textit{Hubble Space Telescope} and various metal elements (including Fe, Ti, Mg, and Na) have been discovered in KELT-9b \citep{Hoeijmakers2018, Cauley2019, Hoeijmakers2019}. 

Theoretically, calcium should exist and probably get ionized into \ion{Ca}{ii} in the upper atmosphere of UHJs.
\cite{Khalafinejad2018} analyzed the \ion{Ca}{ii} near-infrared triplet during the transit of the hot gas giant WASP-17b but did not detect any \ion{Ca}{ii} signals. Very recently, the \ion{Ca}{ii} near infrared triplet lines were detected for the first time in MASCARA-2b, an UHJ with equilibrium temperature $T_\mathrm{eq}$ $\sim$ 2260\,K \citep{Casasayas-Barris2019}.
\ion{Ca}{ii} can also exist in the exospheres of rocky planets \citep{Mura2011}. For example, \ion{Ca}{ii} is detected in the exosphere of Mercury \citep{Vervack2010}.
\cite{Ridden-Harper2016} searched for \ion{Ca}{ii} in the exosphere of the hot rocky planet 55 Cancri e and they found a tentative signal of \ion{Ca}{ii} in one of the four transit datasets. \cite{Guenther2011} attempted to detect calcium in the exosphere of another hot rocky planet, Corot-7b, but were only able to derive an upper limit of the amount of calcium in the exosphere.

Here we report the detections of \ion{Ca}{ii} in two UHJs: KELT-9b and WASP-33b. 
KELT-9b ($T_\mathrm{eq}$ $\sim$ 4050 K) is the hottest exoplanet discovered so far \citep{Gaudi2017} and its host star is a fast-rotating early A-type star. Hydrogen Balmer lines and several kinds of metals \citep{Yan2018, Hoeijmakers2018, Cauley2019}, but not \ion{Ca}{ii}, have been detected in its atmosphere.
WASP-33b ($T_\mathrm{eq}$ $\sim$ 2710 K) is the second hottest giant exoplanet, and it orbits a fast-rotating A5-type star \citep{Cameron2010}. The host star is a $\mathrm{\delta}$ Scuti variable \citep{Herrero2011, Essen2014}.
A temperature inversion, as well as TiO and evidence of AlO, have been detected in the planet \citep{Hayne2015, Nugroho2017, Essen2019}.

The paper is organized as follows. We present the transit observations of the two planets in Section 2. In Section 3, we describe the method to obtain the transmission spectrum of the five \ion{Ca}{ii} lines -- the two H\&K doublet lines and the three near-infrared triplet (IRT) lines. In Section 4, we present the results and discussions including the cross-correlation signal, transmission spectra of individual \ion{Ca}{ii} lines, mixing ratios of \ion{Ca}{i} and \ion{Ca}{ii}, and comparison with models. Conclusions are presented in Section 5.

%
%________________________________________________________________

\section{Observations}
For each of the two planets, we analyzed one transit dataset from CARMENES, which covers the \ion{Ca}{ii} IRT lines and one transit dataset from HARPS-North (HARPS-N), which covers the \ion{Ca}{ii} H\&K lines.

\subsection{WASP-33b observations}
We observed two transits of WASP-33b.
The first transit was observed on 5 January 2017 with the CARMENES  \citep{Quirrenbach2018} spectrograph, installed at the 3.5 m telescope of the Calar Alto Observatory. 
The CARMENES visual channel has a high spectral resolution (R $\sim$ 94\,600) and a wide spectral coverage (520 -- 960\,nm).
A continuous observing sequence of $\sim$ 4.5 hours was performed, covering 0.7 hour before transit and 0.8 hour after transit. The exposure time was set to 120 s, but the first 19 spectra had shorter exposure times (ranging from 65 s to 120 s). The data reduction of the raw spectra was performed with the CARACAL pipeline \citep{Caballero2016}, which includes  bias, flat and cosmic ray corrections, and wavelength calibration. The spectrum produced by the pipeline is at vacuum wavelength and in the Earth's rest frame. We converted the wavelengths into air wavelengths in our study.

The second transit was observed on 8 November 2018 with the HARPS-N spectrograph mounted on the Telescopio Nazionale Galileo. The instrument has a resolution of R $\sim$ 115\,000 and a wavelength coverage of 383 -- 690\,nm. The observation lasted for 9 hours, and we obtained 141 spectra. The raw data were reduced with the HARPS-N pipeline (Data Reduction Software). The pipeline produces order-merged, one-dimensional spectra with a re-sampled wavelength step of 0.01 $\mathrm{\AA}$. The barycentric Earth radial velocity  was already corrected by the pipeline, but we converted it back into the Earth's rest frame in order to be consistent with the CARMENES data. The observation logs of the two transits are summarized in Table~\ref{obs_log}.

\subsection{KELT-9b observations}
We used archival data of one transit from CARMENES observations and one transit from HARPS-N observations. The details of the two observations were described in \cite{Yan2018} and \cite{Hoeijmakers2018}, respectively (see Table \ref{obs_log} for summaries). The CARMENES observation was performed under a partially cloudy weather, thus the spectral signal-to-noise ratio (SNR) was relatively low, and part of the observation was lost due to clouds passing by.

%
%                                             Simple A&A Table
%-----------------------------------------------------------
\begin{table*}
\caption{Observation log.}             % title of Table
\label{obs_log}      % is used to refer this table in the text
\centering                          % used for centering table
\begin{threeparttable}
	\begin{tabular}{l l c c c c c}        % centered columns (4 columns)
	\hline\hline                 % inserts double horizontal lines
~ & Instrument	 & Wavelength coverage &	Date  & Exposure time [s] & $N_\mathrm{spectra}$ \\     % table heading
	\hline      % inserts single horizontal line
WASP-33b & CARMENES	 & 520 -- 960 nm (contains IRT lines) & 2017-01-05	  & 120\tnote{a} & 93 \\ 
 ~ & HARPS-N	 & 383 -- 690 nm (contains H\&K lines) & 2018-11-08	 & 200 & 141 \\      
\hline  
KELT-9b & CARMENES	 & 520 -- 960 nm (contains IRT lines) & 2017-08-06	& 300, 400 & 48 \\ 
~ &	 HARPS-N	 & 383 -- 690 nm (contains H\&K lines) & 2017-07-31	 & 600 & 49 \\      
\hline                                   %inserts single line
	\end{tabular}
	\begin{tablenotes}
        \footnotesize
        \item[a]
        The first 19 spectra had exposure times shorter than 120 s.
     \end{tablenotes}
\end{threeparttable}      
\end{table*}
%-----------------------------------------------------------

%_____________________________________________________________
\section{Method}
We used two different methods to search for and study the \ion{Ca}{ii} lines: 
the cross-correlation method and the direct transmission spectrum of individual lines.
The cross-correlation of high-resolution spectroscopic observations with theoretical model spectra has proven to be a powerful and robust technique to detect molecular and atomic species in exoplanet atmospheres \citep[e.~g.][]{Snellen2010, Alonso-Floriano2019}. 
The direct transmission spectrum method allows detailed study of the line profiles and direct comparison with models \citep[e.~g.][]{Wyttenbach2015, Yan2018}.
In this work, we focused on the \ion{Ca}{ii} H\&K lines (K line 3933.66 $\AA$, H line 3968.47 $\AA$) and the \ion{Ca}{ii} IRT lines (8498.02 $\AA$, 8542.09 $\AA$, 8662.14 $\AA$). These five lines are the strongest \ion{Ca}{ii} lines in the observed wavelength range.

\subsection{Obtaining the transmission spectral matrix}
The transmission spectrum of each \ion{Ca}{ii} line was retrieved separately. We firstly normalized the spectrum and then removed the telluric and stellar lines.

\begin{figure}
   \centering
   \includegraphics[width=0.49\textwidth]{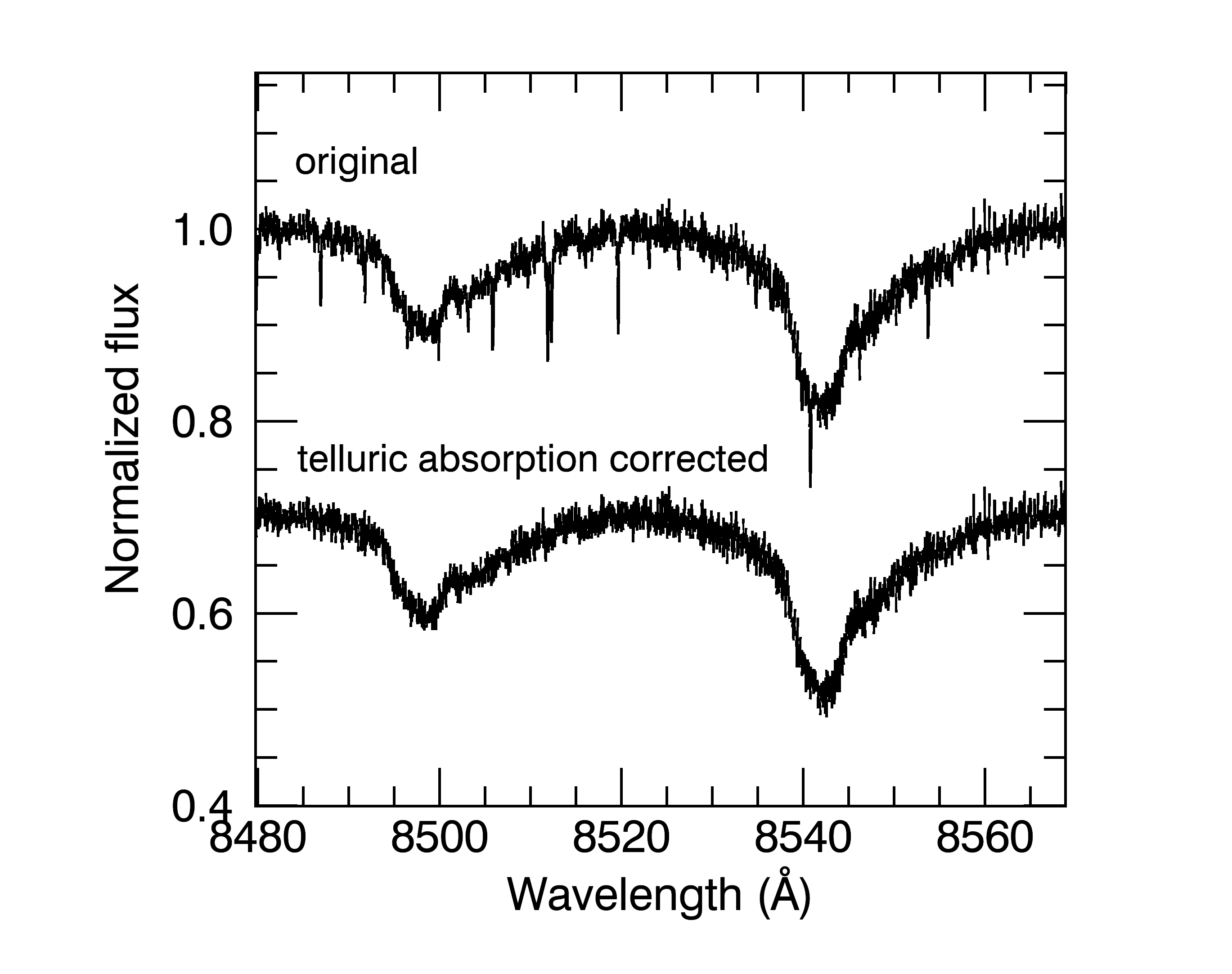}
      \caption{Example of telluric removal. The upper line is the normalized spectrum around two of the \ion{Ca}{ii} IRT lines of KELT-9. The lower line is the telluric absorption corrected spectrum (shifted down by 0.3 for clarity).}
         \label{telluric-remove}
   \end{figure}

The telluric removal was performed in the Earth's rest frame. The telluric lines in the wavelength range of the \ion{Ca}{ii} lines are mostly $\mathrm{H_2O}$ lines. We employed a theoretical $\mathrm{H_2O}$ transmission model described in \cite{Yan2015b}. The model was used to fit and remove the $\mathrm{H_2O}$ lines (Fig. \ref{telluric-remove}).

The removal of the stellar lines was performed by dividing each spectrum with the out-of-transit master spectrum. The master spectrum was obtained by averaging all the observed out-of-transit spectra with their continuum SNRs as weights.
Before obtaining the master spectrum, we aligned all the spectra to the stellar rest frame by correcting the barycentric Earth radial velocity and systemic velocity (--3.0 km\,s$^{-1}$ for WASP-33b and --20.6 km\,s$^{-1}$ for KELT-9b). By performing such a division, the residual spectra during transit should contain the transmission signal from the planetary atmosphere, while the spectra out-of-transit should be normalized to unity.

%                                                Two column figure
%----------------------------------------------------------- S_vib
%-----------------------------------------------------------

%_____________________________________________________________
\subsection{Correction of stellar RM and CLV effects}
During the exoplanet transit, the stellar line profile changes due to the Rossiter-McLaughlin (RM) effect and the center-to-limb variation (CLV) effect. 
The RM effect \citep{Rossiter1924, McLaughlin1924,Queloz2000} causes line profile distortions due to the stellar rotation. 
The CLV effect is the variation of stellar lines across the stellar disk's center to limb \citep{Yan2015a,Czesla2015,Yan2017}, and the effect is mostly the result of differential limb darkening between the stellar line and the adjacent continuum. 

These two effects are encoded in the obtained transmission spectral matrix and the strength of these effects depends on the actual stellar and planetary parameters. The evaluation and correction of the RM and CLV effects are required for transmission spectrum studies. The effects of different lines and different planets have been evaluated, for example, by \cite{Casasayas-Barris2018},  \cite{Yan2018}, \cite{Salz2018}, \cite{Nortmann2018}, and \cite{Keles2019}.

\begin{figure}
   \centering
   \includegraphics[width=0.45\textwidth]{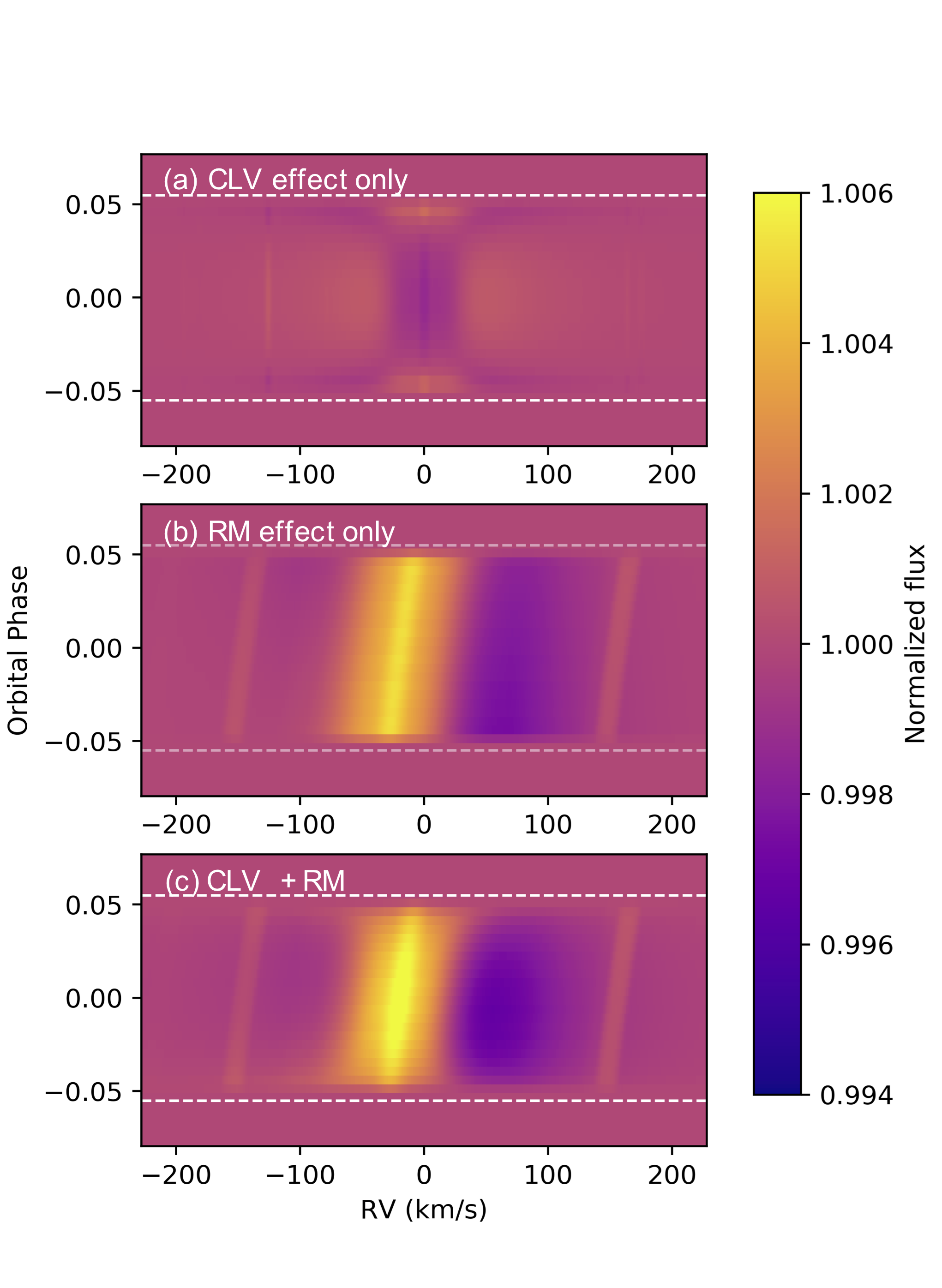}
      \caption{Models of separate CLV and RM effects of the \ion{Ca}{ii} K 3933.66 $\AA$ line for KELT-9b. The upper panel is the CLV effect-only model, the middle panel is the RM effect-only model, and the bottom panel is the model combing both effects.
 For fast rotating stars like KELT-9, the RM effect is stronger than the CLV effect. The simulation here is for the 1 $R_\mathrm{p}$ case.  }    
         \label{Seperate-CLV+RM}
   \end{figure}

We modeled and corrected the RM and CLV effects simultaneously following the method in \cite{Yan2018}. The details of the CLV-only model are described in \cite{Yan2017} and we included the RM effect by assigning the rotational RV to each of the stellar surface elements as done in \cite{Yan2018}.
The RM effect is generally stronger than the CLV effect for these fast rotating stars (c.f. Fig. \ref{Seperate-CLV+RM} for RM-only and CLV-only models).
Actually, the correction of the RM effect is a crucial step in performing the transmission spectroscopy of UHJs because their host stars are normally early-type stars that are fast rotators. 
The stellar and planetary parameters used for the systems of WASP-33b and KELT-9b are listed in Tables \ref{paras_W33} and \ref{paras_K9}, respectively.

The planetary orbit of WASP-33b undergoes a nodal precession as discovered by \cite{Johnson2015}. They measured the changes of orbital inclination and the spin-orbit misalignment angle at two different epochs (2008 and 2014) using Doppler tomography. Measuring the current spin-orbit misalignment during our observations would require additional data reduction procedures such as filtering the stellar pulsation \citep{Johnson2015}, which is beyond the scope of this paper. Therefore, we adopted the change rates and calculated the expected orbital inclination and spin-orbit angle at the dates of our observations. 

As noted by \cite{Yan2018}, the actual effective radius at a given spectral line is larger than 1 $R_\mathrm{p}$ because of the planetary atmospheric absorption. Consequently, the model of the RM + CLV effects should be built with a larger effective radius. We introduced a factor $f$ to account for such an effect by assuming that the actual stellar line profile change is $f$ times the simulated RM + CLV effects with 1 $R_\mathrm{p}$.
In addition, our model has intrinsic errors. For example, our model is in 1D local thermodynamic equilibrium (LTE), while the actual stellar profile is better characterized by a 3D non-LTE stellar model. Thus, such an $f$ factor can also account for the errors of the model. By fitting the observed line profile change with the models using a Markov chain Monte Carlo analysis \citep{Mackey2013}, we obtained $f = 2.1 \pm 0.1$ for the \ion{Ca}{ii} K line and $f = 1.6 \pm 0.2$ for the \ion{Ca}{ii} H line.
For all the other lines, we simply used the 1 $R_\mathrm{p}$ scenario as the data do not have sufficient quality to obtain an $f$ value.

\begin{figure}
   \centering
   \includegraphics[width=0.45\textwidth]{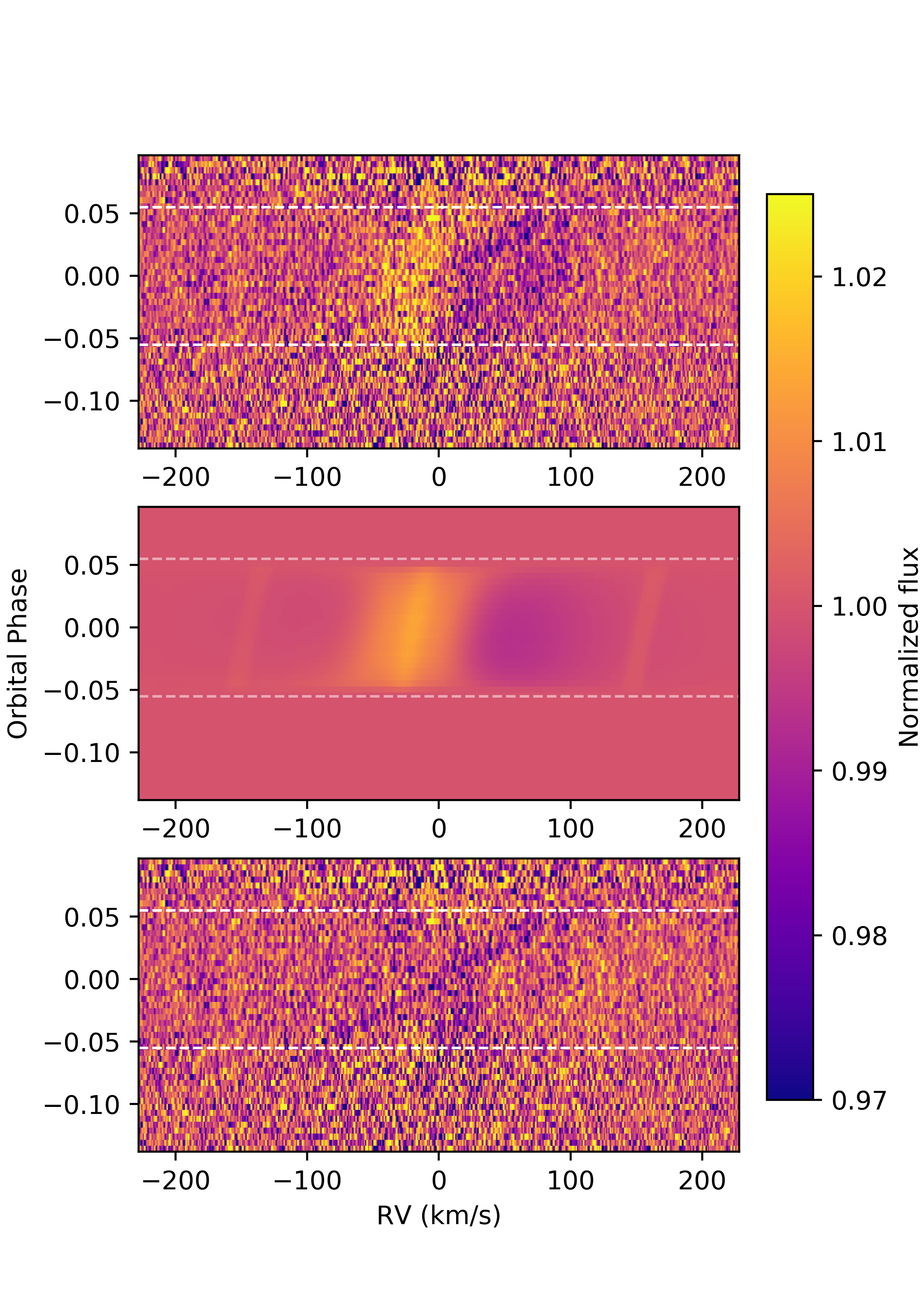}
      \caption{RM + CLV effects of the \ion{Ca}{ii} K 3933.66 $\AA$ line for KELT-9b. \textit{Upper panel:} observed transmission spectral matrix. \textit{Middle panel:} simulated stellar line profile changes due to the RM + CLV effects with an $f$ factor $f = 2.1$ (see text). Separate models of the CLV effect and RM effect are presented in Fig. \ref{Seperate-CLV+RM}.
      \textit{Bottom panel:} transmission spectrum after the correction of the RM + CLV effects. The white horizontal lines label ingress and egress. The obvious shadow with a RV drift from -- 90 km\,s$^{-1}$ at ingress to + 90 km\,s$^{-1}$ at egress is the planetary absorption. }
         \label{CaIIK-SME}
   \end{figure}

Fig.~\ref{CaIIK-SME} presents the model result of the \ion{Ca}{ii} 3933.66 $\AA$ line of KELT-9b as an example. The stellar RM and CLV effects dominate the line profile change. After correcting these stellar effects, we were able to detect the planetary absorption clearly.
The effects of the \ion{Ca}{ii} lines behave differently from the effects of the $\mathrm{H\alpha}$ line \citep[Supplementary Fig.~2 in][]{Yan2018}, demonstrating the importance of modeling the RM and CLV effects for individual lines.

%
%                                             Simple A&A Table
%-----------------------------------------------------------
\begin{table}
\begin{threeparttable}                       % used for centering table
\caption{Parameters of the WASP-33b system.}             % title of Table
\label{paras_W33}      % is used to refer this table in the text
\centering   
\begin{tabular}{l c c  }        % centered columns (4 columns)
\hline\hline                 % inserts double horizontal lines
%~	&	HD189733b &	HD209458b & ~ & ~ \\     % table heading
	Parameter & Symbol [unit] & Value   \\
	\hline                       % inserts single horizontal line
	\textit{The star} & ~ & ~ \\
 		Effective temperature		& $T_\mathrm{eff} [K]$	&	7430 $\pm$ 100\tnote{a}   \\
        Radius 	& $R_\star$ [$R_\odot$]	&	1.509$_{-0.027}^{+0.016}$\tnote{a}    \\
        Mass	& $M_\star$	[$M_\odot$]&	1.561$_{-0.079}^{+0.045}$\tnote{a}    \\
        Metallicity	&  [Fe/H] [dex]	&	--0.1 $\pm$ 0.2\tnote{a}   \\
		Rotational velocity & $v$ sin $i_\star$ [km\,s$^{-1}$]	&	86.63$_{-0.32}^{+0.37}$\tnote{b}   \\	
		Systemic velocity & $v_\mathrm{sys}$ [km\,s$^{-1}$]	&	--3.0 $\pm$ 0.4\tnote{c}   \\		        
        ~ & ~ & ~  \\
        \textit{The planet} & ~ & ~  \\
        Radius  	& $R_\mathrm{p}$ [$R_\mathrm{J}$]	& 1.679$_{-0.030}^{+0.019}$\tnote{a}    \\
        Mass	& $M_\mathrm{p}$	[$M_\mathrm{J}$] 	&	2.16 $\pm$ 0.20\tnote{a}    \\
        Orbital semi-major axis & $a$	[$R_\star$]	 &	3.69 $\pm$ 0.05\tnote{a}   \\
        Orbital period & $P$ [d]	&	1.219870897\tnote{d}  \\
        Transit epoch (BJD) & $T_\mathrm {0}$ [d]	&	2454163.22449\tnote{d}   \\
		Transit depth & $\delta$ [\%] & 1.4\tnote{a} \\		
		RV semi-amplitude & $K_\mathrm{p}$ [km\,s$^{-1}$] & 231 $\pm$ 3\tnote{a}    \\
		Equilibrium temperature & $T_\mathrm{eq}$ [K] &  2710 $\pm$ 50 \\
         ~~ \textit{2017 January 5} & ~ & ~  \\
        Orbital inclination  & $i$ [deg] 	&	89.50\tnote{e}    \\
	    Spin-orbit inclination  & $\lambda$ [deg] 	&	--114.05\tnote{e}   \\
		 ~~ \textit{2018 November 8} & ~ & ~  \\	    
	    Orbital inclination  & $i$ [deg] 	&	90.14\tnote{e}    \\
	    Spin-orbit inclination  & $\lambda$ [deg] 	&	--114.93\tnote{e}    \\

\hline                                   %inserts single line
\end{tabular}
\begin{tablenotes}
        \footnotesize
        \item[a] Adopted from \cite{Lehmann2015} with parameters form \cite{Kovacs2013}.
        \item[b] Adopted from \cite{Johnson2015}.
		\item[c] Adopted from \cite{Nugroho2017}.        
        \item[d] Adopted from \cite{Maciejewski2018}.
        \item[e] Predicted value using parameters in \cite{Johnson2015}.
        
     \end{tablenotes}
\end{threeparttable}      
\end{table}
%-----------------------------------------------------------

%
%                                             Simple A&A Table
%-----------------------------------------------------------
\begin{table}
\begin{threeparttable} 
\caption{Parameters of the KELT-9b system.}             % title of Table
\label{paras_K9}      % is used to refer this table in the text
\centering                         % used for centering table
\begin{tabular}{l c c  }        % centered columns (4 columns)
\hline\hline                 % inserts double horizontal lines
	Parameter & Symbol [unit] & Value\tnote{a}   \\
	\hline                       % inserts single horizontal line
	\textit{The star} & ~ & ~ \\
 		Effective temperature		& $T_\mathrm{eff} [K]$	&	10170 $\pm$ 450   \\
        Radius 	& $R_\star$ [$R_\odot$]	&	2.362$_{-0.063}^{+0.075}$   \\
        Mass	& $M_\star$	[$M_\odot$]& 	2.52$_{-0.20}^{+0.25}$    \\
        Metallicity	&  [Fe/H] [dex]	&	-0.03 $\pm$ 0.2  \\
		Rotational velocity & $v$ sin $i_\star$ [km\,s$^{-1}$]	&	111.4 $\pm$ 1.3   \\	
		Systemic velocity & $v_\mathrm{sys}$ [km\,s$^{-1}$]	&	-20.6 $\pm$ 0.1   \\		        
        ~ & ~ & ~  \\
        \textit{The planet} & ~ & ~  \\
        Radius  	& $R_\mathrm{p}$ [$R_\mathrm{J}$]	& 1.891$_{-0.053}^{+0.061}$    \\
        Mass	& $M_\mathrm{p}$	[$M_\mathrm{J}$] 	&	2.88 $\pm$ 0.84  \\
        Orbital semi-major axis & $a$	[$R_\star$]	 &	3.15 $\pm$ 0.09   \\
        Orbital period & $P$ [d]	&	1.4811235  \\
        Transit epoch (BJD) & $T_\mathrm {0}$ [d]	&	2457095.68572   \\
		Transit depth & $\delta$ [\%] & 0.68 \\		
		RV semi-amplitude & $K_\mathrm{p}$ [km\,s$^{-1}$] & 254$_{-10}^{+12}$     \\
		Equilibrium temperature & $T_\mathrm{eq}$ [K] &  4050 $\pm$ 180 \\
        Orbital inclination  & $i$ [deg] 	&	86.79 $\pm$ 0.25   \\
	    Spin-orbit inclination  & $\lambda$ [deg] 	&	-84.8 $\pm$ 1.4   \\
\hline                                   %inserts single line
\end{tabular}
\begin{tablenotes}
        \footnotesize
        \item[a] All the parameters are adopted from \cite{Gaudi2017}.
     \end{tablenotes}
\end{threeparttable}      
\end{table}
%-----------------------------------------------------------

%                                                Two column figure
%----------------------------------------------------------- S_vib
%-----------------------------------------------------------

%                                                Two column figure
%----------------------------------------------------------- S_vib
%-----------------------------------------------------------

\subsection{Cross-correlation with simulated template}
We simulated the \ion{Ca}{ii} transmission spectra using the \textit{petitRADTRANS} code \citep{Molliere2019}. We assumed that the atmosphere has a solar abundance and an isothermal temperature with a value close to the equilibrium temperature (2700 K for WASP-33b and 4000K for KELT-9b). We also assumed that \ion{Ca}{i} is completely ionized into \ion{Ca}{ii}.
We used a mean molecular weight of 1.3 which is the value for an atomic atmosphere with solar abundance. According to \cite{Hoeijmakers2019}, the transmission spectral continuum due to $\mathrm{H^-}$ absorption in UHJs is normally between 1 mbar and 10 mbar, and the atmosphere below the continuum level cannot be probed. Thus, we set a continuum level of 1 mbar when simulating the transmission spectrum. This was done by adding an absorber of infinite strength for P > 1 mbar. The template spectra were subsequently convolved with the instrument profiles. We established a grid of template spectra with radial velocity (RV) shifts from -- 500 km\,s$^{-1}$ to + 500 km\,s$^{-1}$ with a step of 1 km\,s$^{-1}$. 

Before cross-correlation, we filtered the residual spectra using a Gaussian filter with a $\sigma$ of $\sim$ 1.5 $\AA$. In this way, we filtered out large scale spectral features, which could be caused by the blaze function variation and the stellar pulsation. We cross-correlated the residual spectra with the simulated template spectra as in \cite{Snellen2010}.
The cross-correlations of the \ion{Ca}{ii} H\&K from HARPS-N observations and the \ion{Ca}{ii} IRT from CARMENES observations were performed independently. For each observed spectrum, we obtained one cross-correlation function (CCF; Fig.~\ref{map-W33}b).
We then combined all the in transit CCFs by shifting them to the planetary rest frame for a given $K_\mathrm{p}$ (RV semi-amplitude of planetary orbital motion). In this way, we generated the $K_\mathrm{p}$-map with $K_\mathrm{p}$ ranging from 0 to 400 km\,s$^{-1}$ with a step of 1 km\,s$^{-1}$ (Fig.~\ref{map-W33}c). This two-dimensional CCF map has been widely used in previous cross-correlation studies, as in Figure 8 of \cite{Birkby2017} and Figure 14 of \cite{Nugroho2017}.
In order to estimate the SNR, we measured the noise of the $K_\mathrm{p}$ map as the standard deviation of CCF values with RV ranges of --200 to --100 km\,s$^{-1}$ and +100 to +200 km\,s$^{-1}$. Each $K_\mathrm{p}$-map was then divided by the corresponding noise value.

\section{Results and discussion}

\subsection{Detection of \ion{Ca}{ii} using the cross-correlation method}
\begin{figure*}
   \centering
   \includegraphics[width=0.95\textwidth]{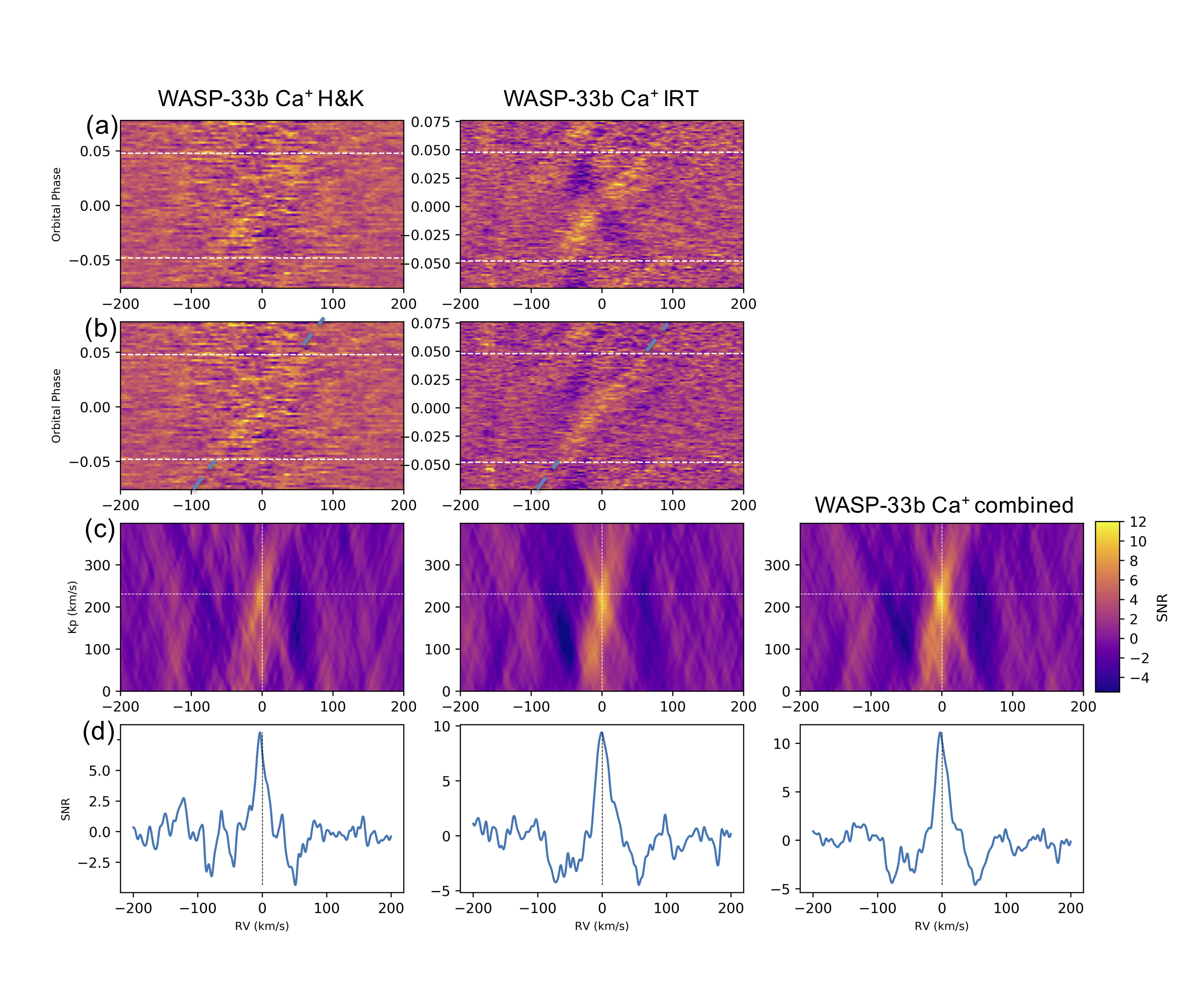}
      \caption{Cross-correlation results of WASP-33b for \ion{Ca}{ii} H\&K (\textit{left panels}), \ion{Ca}{ii} IRT \textit{(middle panels)} and the combined five lines \textit{(right panels)}.
\textit{(a)} The CCF maps without the correction of stellar effects. The two horizontal dashed lines indicate the time of ingress and egress. 
\textit{(b)} The CCF maps with stellar RM+CLV effects corrected. The correction of RM+CLV effects were performed on the transmission spectral matrix before the cross-correlation. There is a remaining black stripe at --30 km\,s$^{-1}$ in the CCF map of \ion{Ca}{ii} IRT. It is probably the feature of stellar pulsation because WASP-33 is a variable star.
\textit{(c)} The $K_\mathrm{p}$-map. These are the combined in-transit CCFs for different $K_\mathrm{p}$ values. The horizontal dashed line marks the expected $K_\mathrm{p}$ calculated using the planetary orbital parameters from the literature. The vertical dashed line marks RV = 0 km\,s$^{-1}$, where the planetary signal is expected to be located (we have corrected the stellar systemic RV already).
\textit{(d)} The CCF at $K_\mathrm{p}$ = 231 km\,s$^{-1}$ (expected $K_\mathrm{p}$ value).
}
         \label{map-W33}
   \end{figure*}

\begin{figure*}
   \centering
   \includegraphics[width=0.95\textwidth]{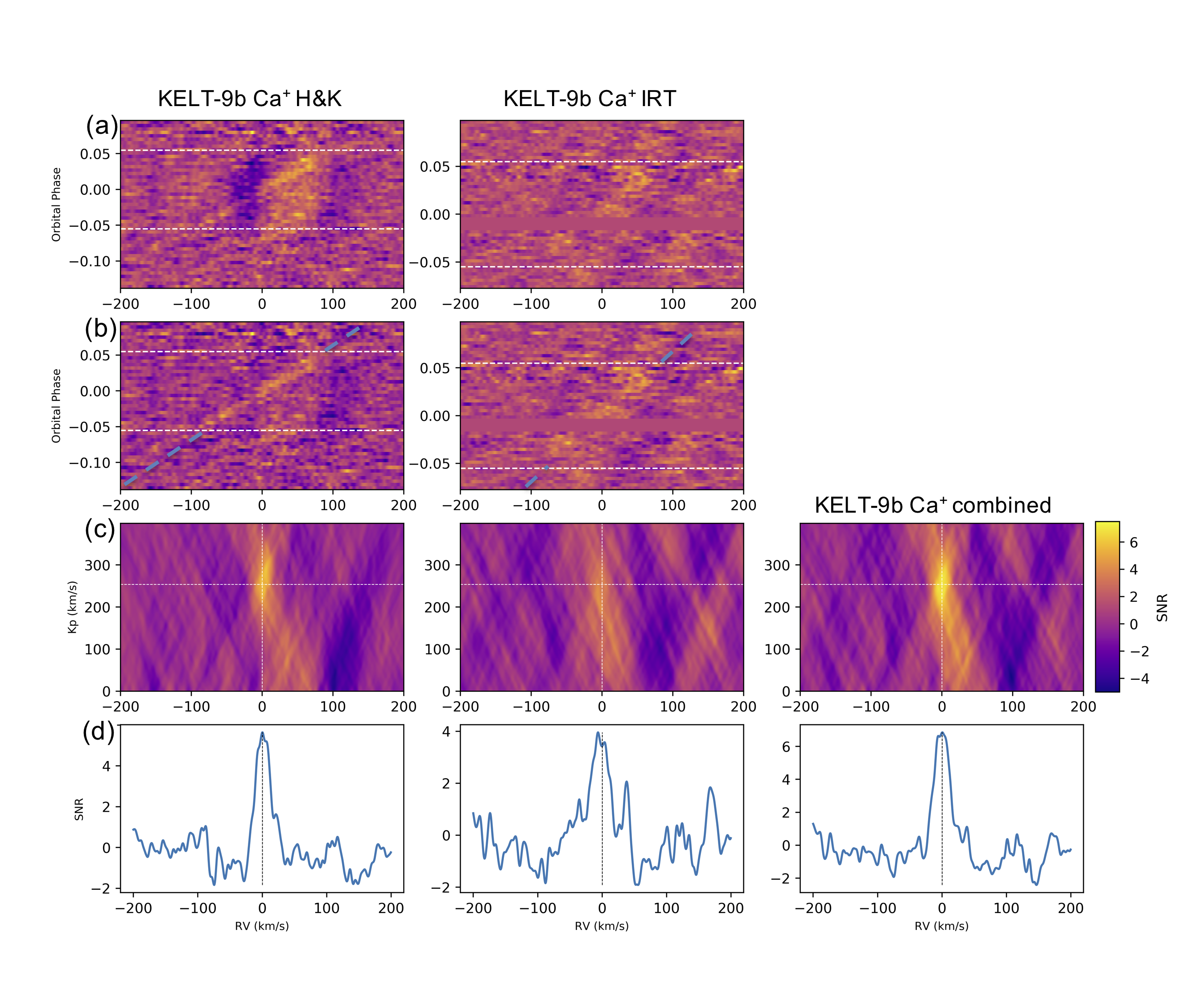}
      \caption{Same as Fig.~\ref{map-W33} but for KELT-9b. The expected $K_\mathrm{p}$ value for KELT-9b is 254 km\,s$^{-1}$.}
         \label{map-K9}
   \end{figure*}

The cross-correlation results of the two planets are presented in Figs.~\ref{map-W33} and \ref{map-K9}. Panel \textit{b} in the figures are cross-correlation maps between the model spectrum and the residual spectrum with the RM+CLV effects corrected. We also calculated the cross-correlation maps between the model spectrum and the original residual spectrum for comparison (panel \textit{a}).

The \ion{Ca}{ii} H\&K and the \ion{Ca}{ii} IRT lines are detected in both planets. We further added the $K_\mathrm{p}$-maps of H\&K and IRT to obtain the combined $K_\mathrm{p}$-map of the five \ion{Ca}{ii} lines (right panel in the figures). Here we added directly the H\&K and IRT $K_\mathrm{p}$-maps divided by their corresponding noise values.
The combined $K_\mathrm{p}$ maps show strong cross-correlation signals at the expected $K_\mathrm{p}$ values ($231\pm3$ km\,s$^{-1}$ for WASP-33b and 254$_{-10}^{+12}$ km\,s$^{-1}$ for KELT-9b). These $K_\mathrm{p}$ values are calculated using Kepler's third law with orbital parameters from the literature (Tables \ref{paras_W33} and \ref{paras_K9}). The peak SNR value in the $K_\mathrm{p}$-map is located at $K_\mathrm{p}$=224 km\,s$^{-1}$ for WASP-33b and $K_\mathrm{p}$=266 km\,s$^{-1}$ for KELT-9b. For KELT-9b, \cite{Yan2018} derived $K_\mathrm{p}$=$269\pm6$ km\,s$^{-1}$ using H$\mathrm{\alpha}$ absorption and \cite{Hoeijmakers2019} obtained a $K_\mathrm{p}$ value of $234.2\pm0.9$ km\,s$^{-1}$ using the planetary \ion{Fe}{II} absorption lines. These $K_\mathrm{p}$ values derived from planetary absorption are different, but broadly consistent, with the expected $K_\mathrm{p}$ values derived from orbital parameters. Considering that the planetary atmosphere may have additional RV components originated from dynamics, we decided to use the expected $K_\mathrm{p}$ values from Kepler's third law in the rest of the paper.

The bottom panel in the figures presents the CCFs at the expected $K_\mathrm{p}$ values.
Since we already corrected for the systemic velocity as described in Section 3.1, the planetary signal is expected to be located at RV $\sim$ 0 km\,s$^{-1}$. 

For WASP-33b,  the \ion{Ca}{ii} IRT  signal is very clear and can be seen in the CCF map directly (middle panel in Fig.~\ref{map-W33}b). The \ion{Ca}{ii} H\&K signal is also strong but is less significant than the IRT lines, probably because the deep stellar \ion{Ca}{ii} H\&K lines significantly reduce the flux level.
The combined cross-correlation function of the five lines yields a 11 $\sigma$ detection.

For KELT-9b, the CCF map of the \ion{Ca}{ii} H\&K lines shows a clear signal.
However, the \ion{Ca}{ii} IRT signal is less significant, which we attribute to the bad weather conditions during the CARMENES observation. Nevertheless, the IRT signal is still detected at a 4 $\sigma$ level as shown in Fig.~\ref{map-K9}d.
The combined CCF shows a 7 $\sigma$ detection.

The correction of the stellar RM and CLV effects plays an important role in the \ion{Ca}{ii} detection. 
By comparing the CCF maps with and without the stellar correction in Fig.~\ref{map-W33} and Fig.~\ref{map-K9}, the improvement after the correction is significant. \cite{Hoeijmakers2019} searched for \ion{Ca}{ii} in KELT-9b using the same HARPS-N data as in this work, however, they were not able to detect the \ion{Ca}{ii} H\&K lines. 
Probably, the different treatment of the RM and CLV effects between our works could be responsible for the different results obtained.
They used an empirical model to fit the stellar residuals presented in the CCF map, which did not properly correct the stellar RM and CLV effects of the H\&K lines (see Fig.~\ref{CaIIK-SME}).

Stellar chromospheric activity can potentially affect the planetary \ion{Ca}{ii} detection but is not expected to pose a serious problem in early A-type stars \citep[e.~g.][]{Schmitt1997}. \cite{Cauley2018} and \cite{Khalafinejad2018} investigated the effect of stellar activity on transmission spectra of calcium lines. 
One prominent distinction between stellar activity and planetary absorption is the radial velocity difference \citep{Barnes2016}. In our work, the detected \ion{Ca}{ii} signals follow the expected orbital velocities of KELT-9b and WASP-33b, which strongly support their planetary origin.

The host star of WASP-33b is a variable star and stellar pulsations could potentially change the stellar line profile \citep{Cameron2010}. However, since the \ion{Ca}{ii} feature occurs at the planetary velocity and only during transit (c.~f. the middle figure in Fig.~\ref{map-W33}b), the detected signal is unlikely to be the result of stellar pulsation.  Although the planetary atmosphere feature is unambiguous, the stellar pulsation features probably affect the CCF map. For instance, there is a feature in the IRT CCF map at around --30 km\,s$^{-1}$ (see middle panel in Fig.~\ref{map-W33}b). Such a pulsation feature produces a dark region next to the planetary atmosphere signal on the $K_\mathrm{p}$-map (middle panel in Fig.~\ref{map-W33}c).

%                                                Two column figure
%----------------------------------------------------------- S_vib
%-----------------------------------------------------------

%                                                Two column figure
%----------------------------------------------------------- S_vib
%-----------------------------------------------------------

%                                                Two column figure
%----------------------------------------------------------- S_vib
   \begin{figure*}
   \centering
   \includegraphics[width=0.95\textwidth]{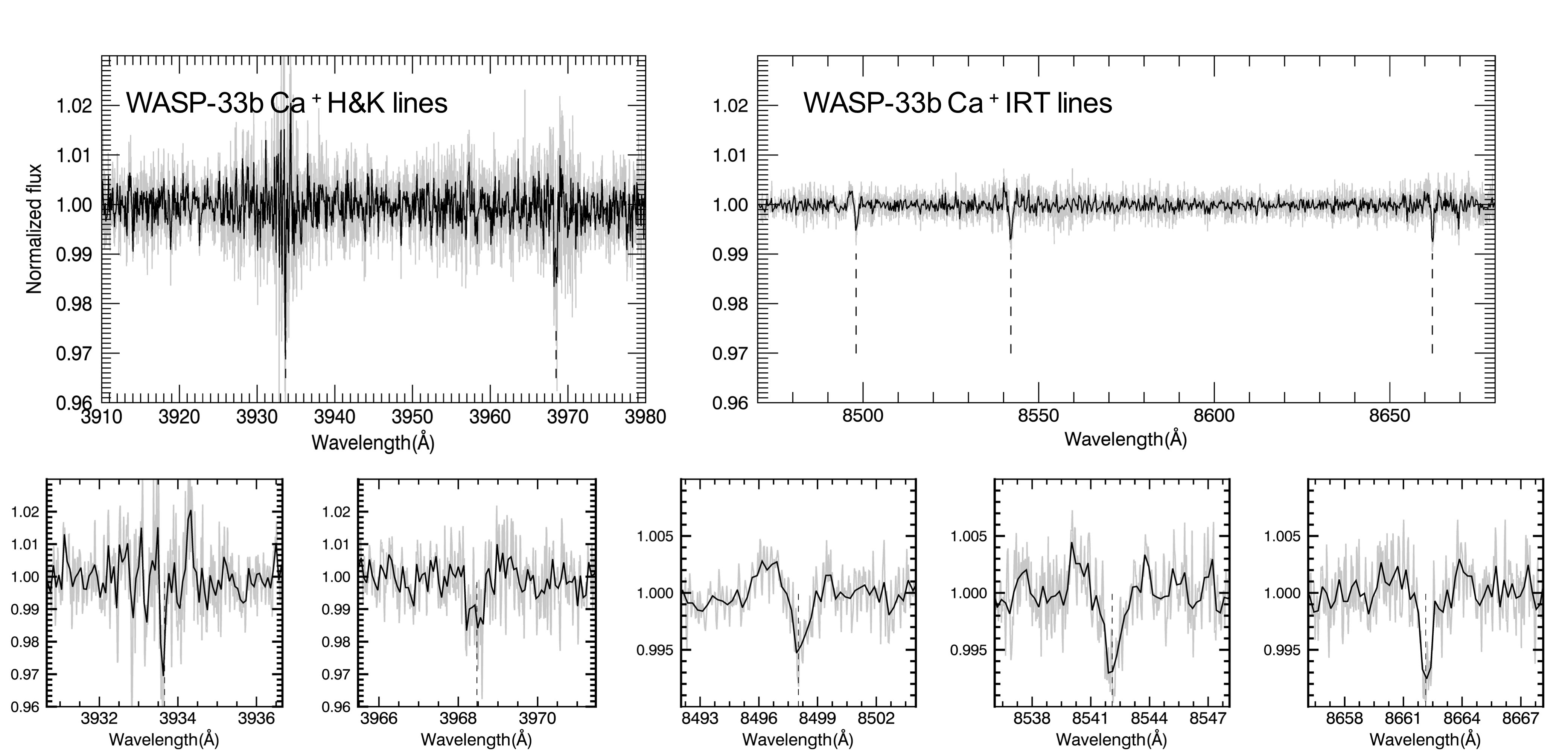}
      \caption{Transmission spectra of the \ion{Ca}{ii} lines for WASP-33b. They were obtained by combining all the in-transit spectra (excluding ingress and egress). Grey lines denote the original spectra and black lines the binned spectra (7 points bin). Dashed vertical lines indicate the rest wavelengths of the line centers. \textit{Upper panels:} the \ion{Ca}{ii} H\&K and IRT lines plotted with the same y-axis scale. The strength of the H\&K lines is significantly stronger than that of the IRT lines. \textit{Lower panels:} enlarged view of each of the five lines.}
         \label{tran-W33}
   \end{figure*}
%-----------------------------------------------------------

%                                                Two column figure
%----------------------------------------------------------- S_vib
   \begin{figure*}
   \centering
   \includegraphics[width=0.95\textwidth]{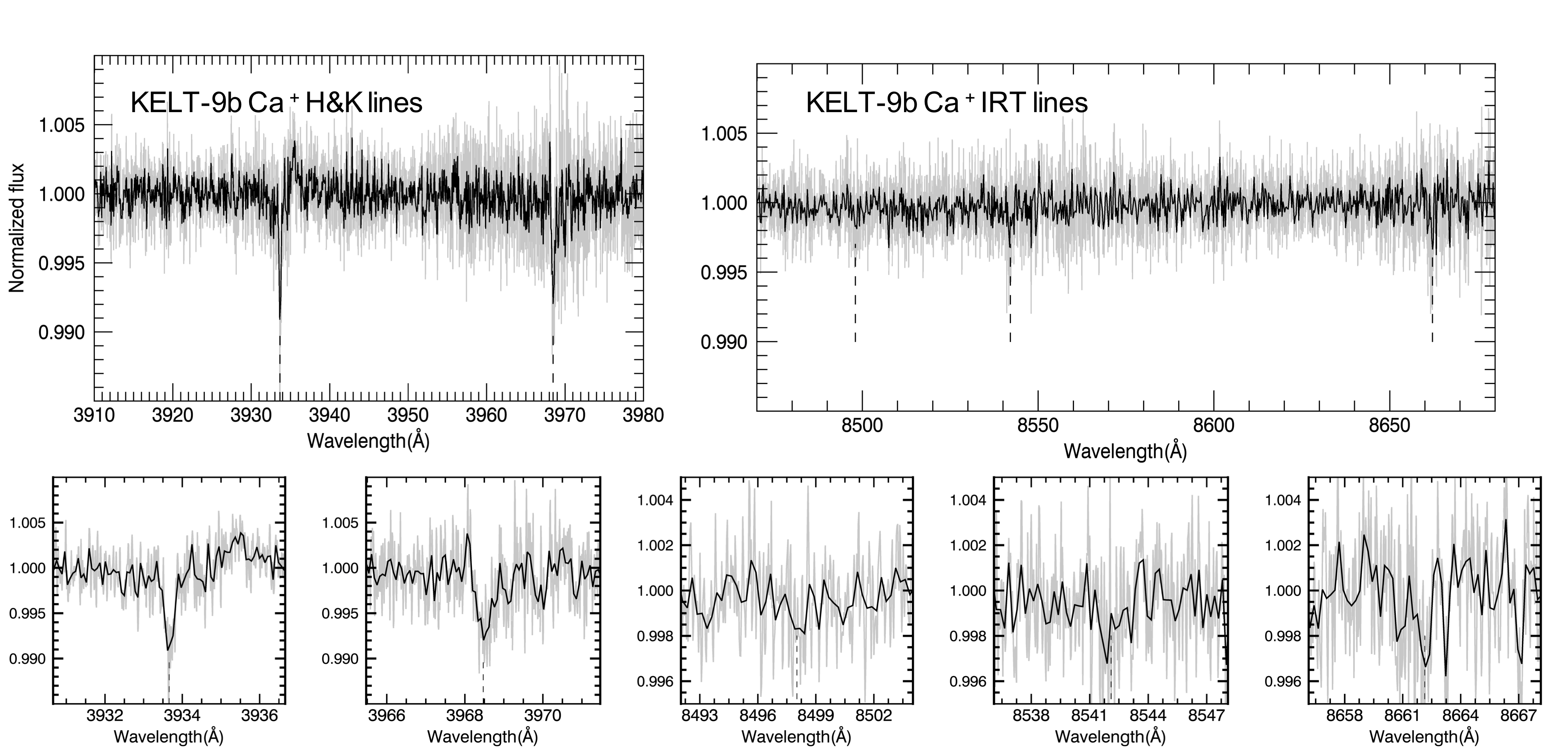}
      \caption{Same as Fig.~\ref{tran-W33} but for KELT-9b.}
         \label{tran-K9}
   \end{figure*}
%-----------------------------------------------------------

\subsection{Transmission spectrum of individual \ion{Ca}{ii} lines}
In addition to the \ion{Ca}{ii} detection using the cross-correlation technique, we also obtained the transmission spectrum of each individual line in order to study them in detail (Figs.~\ref{tran-W33} and \ref{tran-K9}). We added up all the residual spectra observed in transit (but excluding the ingress and egress). Here the stellar RM and CLV effects are already corrected in the residual spectra. Before adding up, we shifted the spectra to the planetary rest frame using the literature $K_\mathrm{p}$ values. 

\begin{figure*}
   \centering
   \includegraphics[width=0.88\textwidth]{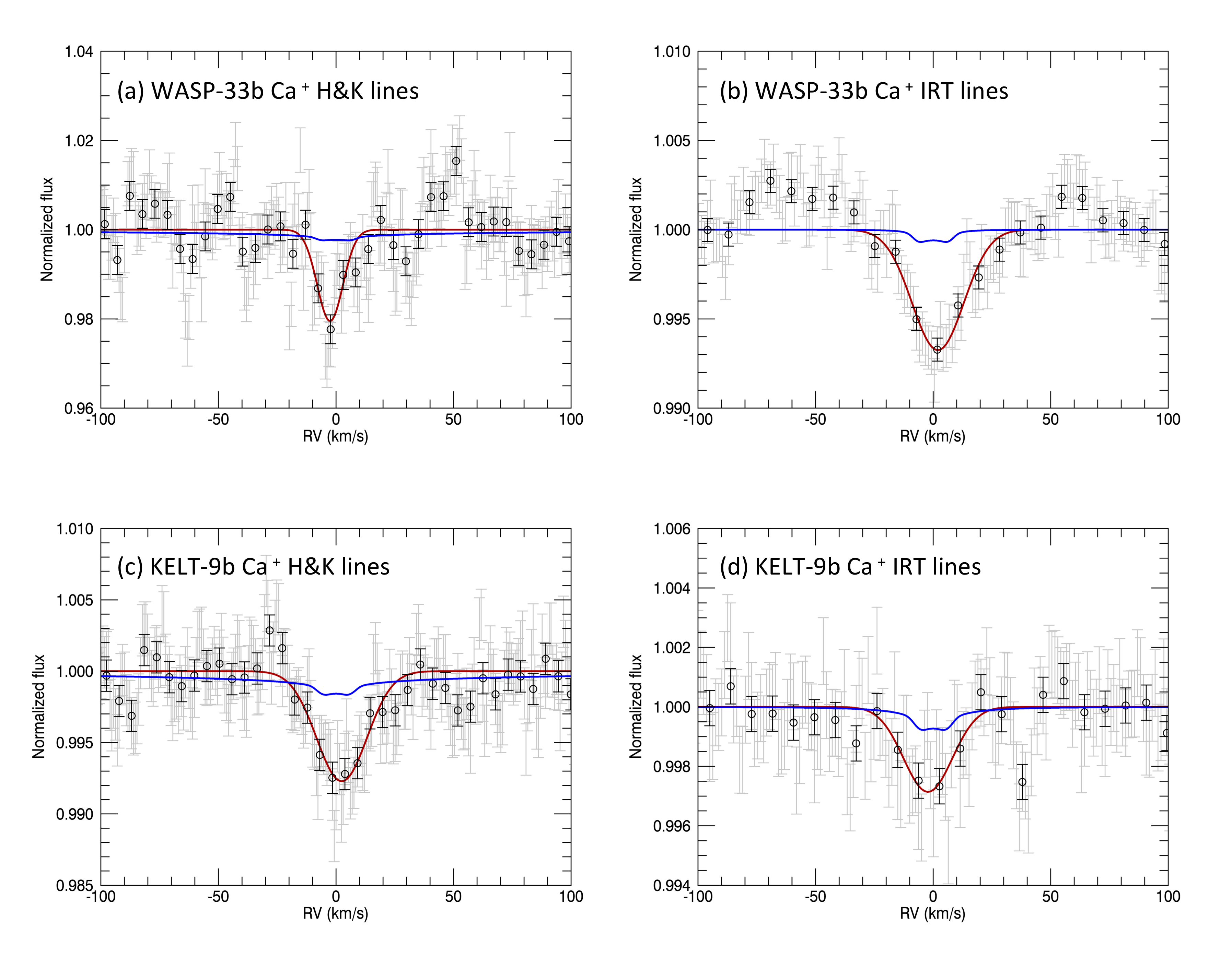}
      \caption{Average line profiles of \ion{Ca}{ii} H\&K (\textit{left}) and \ion{Ca}{ii} IRT (\textit{right}) for WASP-33b (\textit{top}) and KELT-9b (\textit{bottom}). The grey points are original transmission spectra and the black circles are spectra binned every 7 points. The red lines are Gaussian fits to the line profiles. The fitted results are presented in Table \ref{fit_result}. The blue lines are model spectra calculated assuming isothermal temperatures and tidal locked rotation.}
         \label{fig-profiles}
   \end{figure*}

In order to study the absorption line profile, we averaged the H\&K lines as well as the triplet lines (see Fig.~\ref{fig-profiles}). 
We performed the averaging based on the facts that the line depths of the two H\&K lines are similar and the line depths of the triplet lines are also similar.
Subsequently, a Gaussian function is fitted to the average spectrum. The fit results are presented in Table \ref{fit_result}. 
We measured the standard deviation of the spectrum in the ranges of --200 to --100 km\,s$^{-1}$ and +100 to +200 km\,s$^{-1}$ and assigned this value as the error of each data point.

The average line depth of the H\&K lines is significantly larger than that of the IRT lines. The line depth ratio between them is $3.0\pm0.3$ for WASP-33b and $2.7\pm0.5$ for KELT-9b. This is because the H\&K lines correspond to resonant transitions from the ground state of \ion{Ca}{ii}, while the IRT lines are not. For both planets, the 8542 $\AA$ line is the strongest and the 8498  $\AA$ line is the weakest among the three IRT lines. These relative line strengths are consistent with the transmission spectral model.

We calculated the effective radius at the line center ($R_\mathrm{eff}$) and compared it with the effective Roche lobe radius at the planetary terminator \citep{Ehrenreich2010}.
$R_\mathrm{eff}$ is obtained using the equation: $\pi R_\mathrm{eff}^2 / \pi R_\mathrm{p}^2 = (\delta + h) / \delta$, where $\delta$ is the optical photometric transit depth (c.~f. Tables~\ref{paras_W33} and \ref{paras_K9}) and $h$ is the observed line depth.
For WASP-33b, the $R_\mathrm{eff}$ of H\&K lines is 1.56 $R_\mathrm{p}$ which is very close to the effective Roche radius ($1.71_{-0.07}^{+0.08}~R_\mathrm{p}$). For KELT-9b, the $R_\mathrm{eff}$ of H\&K lines is 1.47 $R_\mathrm{p}$ and the effective Roche radius is $1.91_{-0.26}^{+0.22}~R_\mathrm{p}$. 
Therefore, the effective radii of both planets are close but below the Roche radii. 
As a result, we infer that the ionized calcium detected here mostly originates from the extended atmospheric envelope within the Roche lobe instead of from the already escaped material beyond the Roche lobe. Escaped material beyond the Roche lobe can potentially form a comet-like tail as detected in some exoplanets using the hydrogen Lyman-$\mathrm{\alpha}$ and the helium 1083 nm absorptions \citep[e.~g.][]{Ehrenreich2015,Nortmann2018}.

The full width at half maximum (FWHM) of the average line profile is also presented in Table \ref{fit_result}.
The FWHM values of the \ion{Ca}{ii} lines in KELT-9b agree in general with the values of other metal lines in KELT-9b measured by \cite{Cauley2019} and \cite{Hoeijmakers2019}. The observed line width is probably a combined result of thermal broadening, rotational broadening, and hydrodynamic escape velocity.

The measured line centers have blue- or red-shifted RVs of several km\,s$^{-1}$ ($v_\mathrm{center}$ values in Table \ref{fit_result}). However, 
we do not claim the detection of atmospheric winds considering the large errors. The residuals of the RM and CLV effects as well as stellar pulsation can potentially affect the obtained transmission line profile.
Furthermore, the uncertainty of the stellar systemic velocity also affects the measurement of $v_\mathrm{center}$. For example, \cite{Gaudi2017} reported $v_\mathrm{sys}$ of KELT-9 to be --20.6 $\pm$ 0.1 km\,s$^{-1}$. However, \cite{Hoeijmakers2019} obtained a value of --17.7 $\pm$ 0.1 km\,s$^{-1}$ using HARPS-N observations. Later, \cite{Borsa2019} measured $v_\mathrm{sys}$ of KELT-9 as --19.819 $\pm$ 0.024 km\,s$^{-1}$ also using HARPS-N data. For WASP-33, \cite{Nugroho2017} found that their measured $v_\mathrm{sys}$ deviates from the values in other works by $\sim$ 1\,km\,s$^{-1}$. 
The discrepancy between the $v_\mathrm{sys}$ measurements could be due to different instrument RV zero-points,  stellar templates, and methods to measure RV, as well as to the relatively large stellar variability of WASP-33. In principle, radial velocity measurement of fast rotating early type stars like WASP-33 and KELT-9 is intrinsically difficult because of the lack of sufficient stellar lines and the broad line profile.
Therefore, one should be cautious when interpreting the measured shift of the line center as the signature of atmospheric winds.

\begin{figure*}
   \centering
   \includegraphics[width=0.95\textwidth]{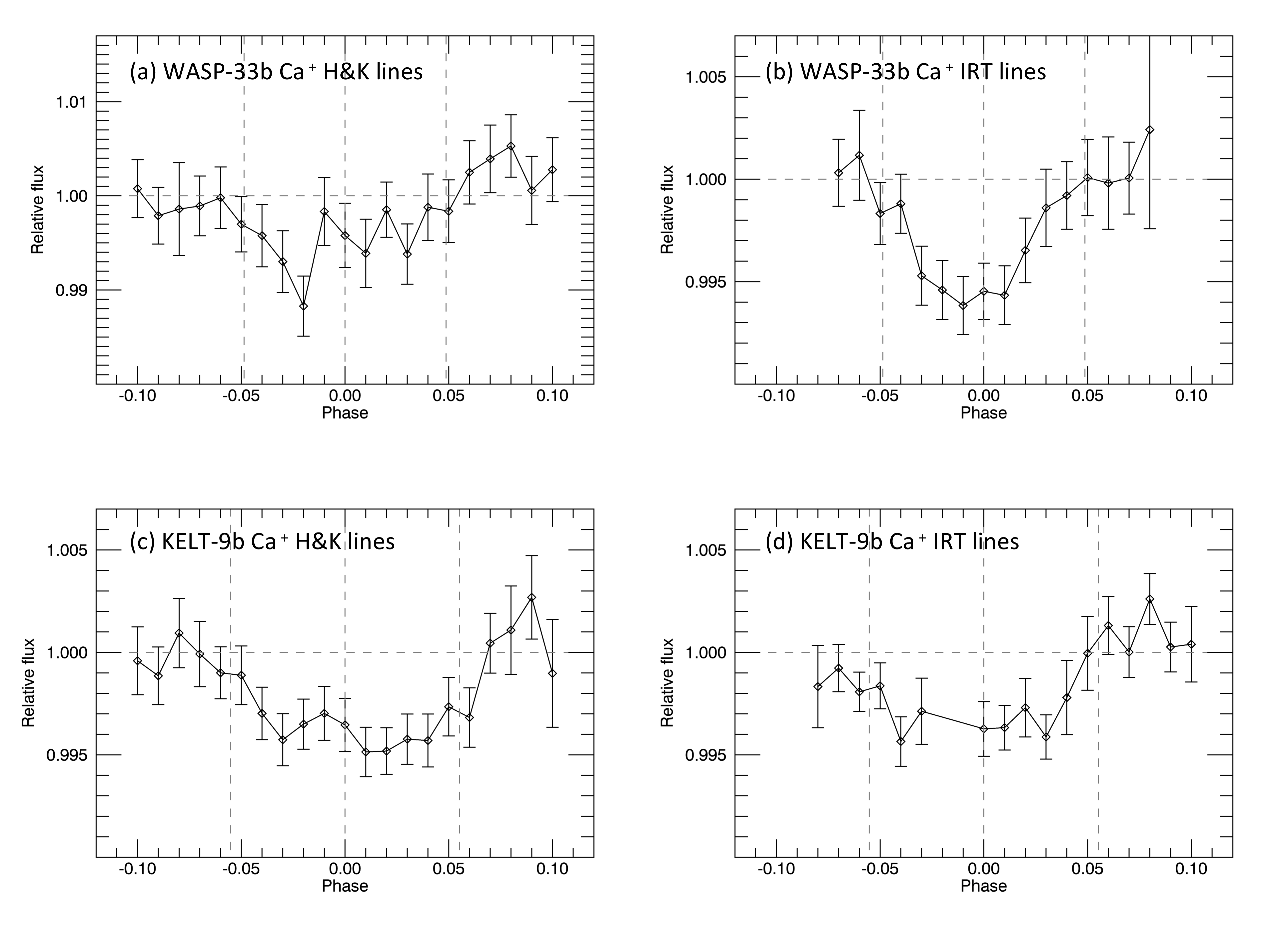}
      \caption{Average light curves of the \ion{Ca}{ii} H\&K lines (\textit{left panels}) and \ion{Ca}{ii} IRT lines (\textit{right panels}). These are relative fluxes measured with an 1 $\AA$ band centered at the line core. The measurement was performed in the planetary rest frame and the stellar RM+CLV effects were corrected.}
         \label{fig-LC}
   \end{figure*}

In order to investigate the time series of the \ion{Ca}{ii} absorption, we measured the relative fluxes of each \ion{Ca}{ii} line in the residual spectra with an 1 $\AA$ band centered at the line core. Subsequently, we averaged the obtained light curves of the H\&K lines and the IRT lines and binned the data points with a phase step of 0.01 (Fig.~\ref{fig-LC}).  
The light curves of the WASP-33b IRT lines and the KELT-9b H\&K lines have clear absorption signals during transit. However, the absorption signals are less prominent in the light curves of the WASP-33b H\&K lines and the KELT-9b IRT lines,  which is probably due to lower data quality, stellar pulsation noise, and residuals of the RM + CLV effects. 

%                                                Two column figure
%----------------------------------------------------------- S_vib
%-----------------------------------------------------------

%                                                Two column figure
%----------------------------------------------------------- S_vib
%-----------------------------------------------------------

%
%                                             Simple A&A Table
%-----------------------------------------------------------
\begin{table*}
\caption{Line profile parameters obtained by fitting a Gaussian function to the average line profile in Fig.~\ref{fig-profiles}. }             % title of Table
\label{fit_result}      % is used to refer this table in the text
\centering                          % used for centering table
\begin{threeparttable}
	\begin{tabular}{c c c c c c c}        % centered columns (4 columns)
	\hline\hline                 % inserts double horizontal lines
~ & Lines	 & $v_\mathrm{center}$ [km\,s$^{-1}$] & FWHM [km\,s$^{-1}$]  & Line depth  & Model line depth & $R_\mathrm{eff}$ [$R_\mathrm{p}$] \\     % table heading %& 
	\hline      % inserts single horizontal line

\textbf{WASP-33b} & \ion{Ca}{ii} H\&K & --1.9 $\pm$ 0.7 & 15.2 $\pm$ 1.5	  & (2.02 $\pm$ 0.17)$\%$  & 0.235$\%$ & 1.56 $\pm$ 0.04 \\      
 ~ & \ion{Ca}{ii} IRT	 & 2.0 $\pm$ 0.7 & 26.0 $\pm$ 1.7	  & (0.67 $\pm$ 0.04)$\%$  & 0.064$\%$ & 1.22 $\pm$ 0.01 \\ 
 
\hline  
\textbf{KELT-9b} & \ion{Ca}{ii} H\&K	 &  3.2 $\pm$ 0.7 & 27.5 $\pm$ 1.8	& (0.78 $\pm$ 0.04)$\%$ & 0.162$\%$ & 1.47 $\pm$ 0.02 \\
~ &	\ion{Ca}{ii} IRT	& --2.2 $\pm$ 2.2 & 24.4 $\pm$ 5.1	& (0.29 $\pm$ 0.05)$\%$  & 0.076$\%$ & 1.19 $\pm$ 0.03 \\      
\hline                                   %inserts single line
	\end{tabular}
	\begin{tablenotes}
      \small
      \item \textbf{Notes.} Here $v_\mathrm{center}$ is the measured RV shift of the line center compared to the theoretical line center. The model line depth is the average line depth from the \ion{Ca}{ii} transmission model. $R_\mathrm{eff}$ is the effective radius at the line center calculated using the observed line depth value and the photometric transit depth. 
    \end{tablenotes}
\end{threeparttable}      
\end{table*}
%-----------------------------------------------------------

\subsection{Mixing ratios of \ion{Ca}{i} and \ion{Ca}{ii}}
\begin{figure}
   \centering
   \includegraphics[width=0.45\textwidth]{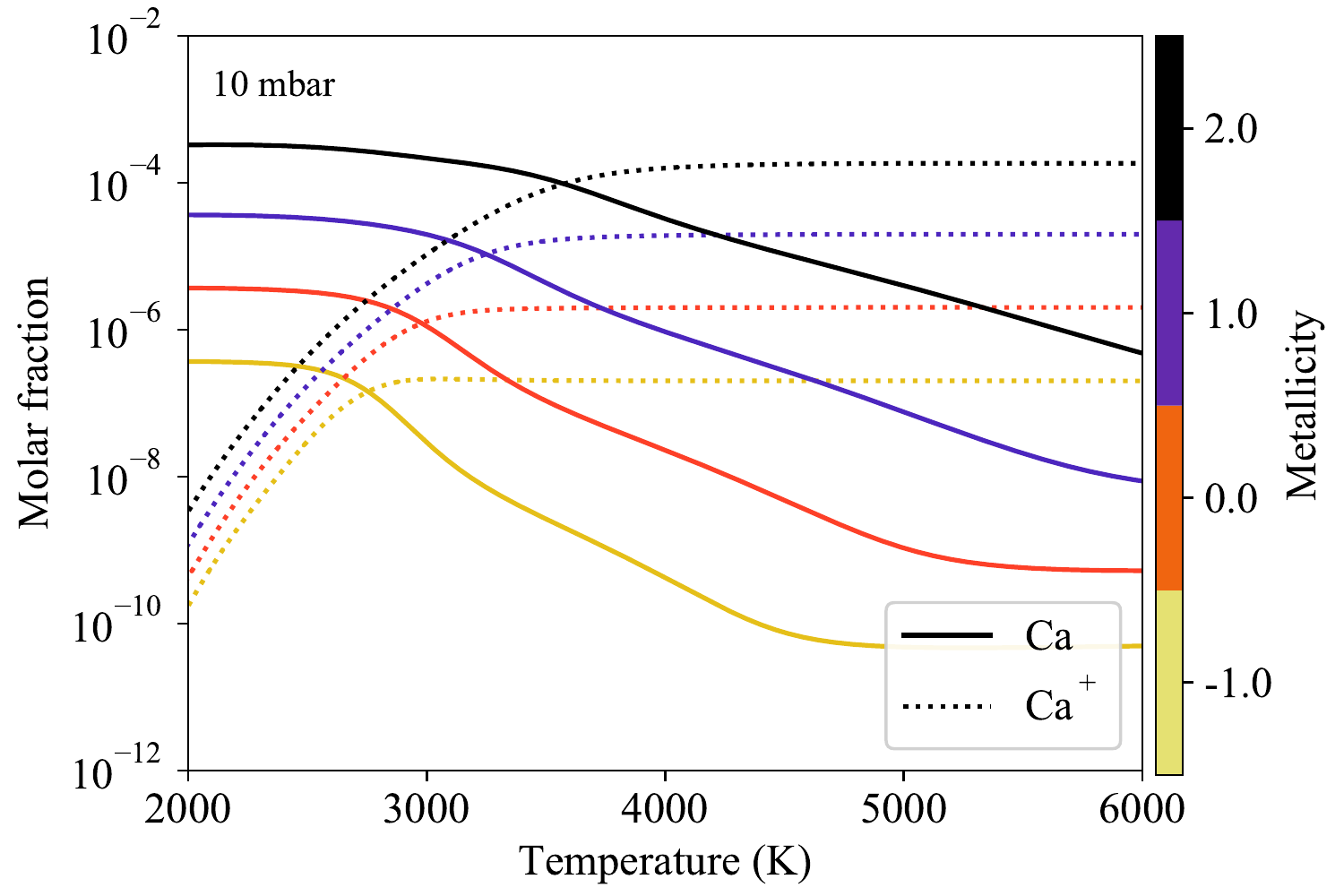}
      \caption{The \ion{Ca}{i} (solid line) and \ion{Ca}{ii} (dashed line) mixing ratios as a function of temperature. Here we assumed chemical equilibrium without photo-ionization and a pressure of 10 mbar. Different colors label different metallicities ([Fe/H] in unit of dex).
      }
         \label{Ca-temperature}
   \end{figure}

We calculated the equilibrium chemistry between atomic calcium (\ion{Ca}{i}) and singly ionized calcium (\ion{Ca}{ii}) using the chemical module in the \textit{petitCode} \citep{Molliere2015, Molliere2017}. Figure \ref{Ca-temperature} shows the mixing ratio variation with temperature. For an atmosphere with a solar metallicity and at a pressure of 10 mbar, ionized calcium becomes dominant at temperatures higher than 3000 K. In general, the calcium ionization rate is higher with higher temperature and lower pressure.

As calculated by \cite{Hoeijmakers2019}, the spectral continuum level for UHJ is typically located between 1~mbar and 10~mbar, and transmission spectroscopy probes only the atmosphere above the continuum level. Therefore, we expect \ion{Ca}{ii} to be the dominant calcium feature in the transmission spectroscopy of UHJs. \ion{Ca}{i} can be probed at lower altitudes if the planetary atmosphere is cooler than 3000 K. For example, atomic calcium was detected in the relatively cool hot-Jupiter HD 209458b, which has a $T_\mathrm{{eq}}$ of 1460 K \citep{Astudillo-Defru2013}.
Here we did not include photo-ionization in the chemical model. The ionization fraction will be higher when including photo-ionization.

\begin{figure}
   \centering
   \includegraphics[width=0.45\textwidth]{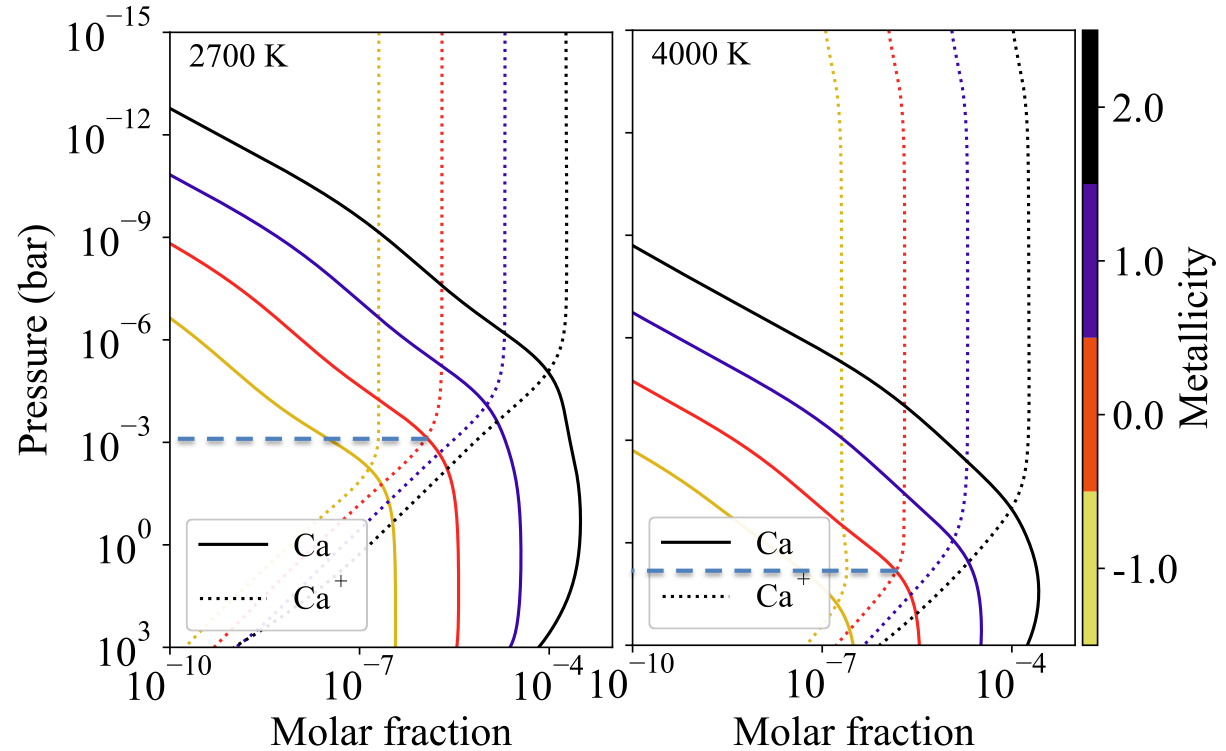}
      \caption{Neutral and ionized calcium profiles for WASP-33b (\textit{left}) and KELT-9b (\textit{right}). Here we assumed isothermal temperatures as indicated in the top left corner of each panel. At solar metallicity (orange lines), \ion{Ca}{ii} is dominant at high altitudes with pressure < 1 mbar for WASP-33b and with pressure < 10~bar for KELT-9b (denoted with horizontal dashed lines).
      }
         \label{Mixing-ratio}
   \end{figure}

Figure \ref{Mixing-ratio} shows the \ion{Ca}{i} and \ion{Ca}{ii} mixing ratios of the two planets assuming isothermal temperature distributions. For WASP-33b, \ion{Ca}{ii} is dominant at high altitudes with pressures < 1 mbar assuming solar metallicity; for KELT-9b, \ion{Ca}{ii} is dominant at pressures < 10 bar. According to our simulations, the average $\mathrm{H^-}$ continuum level is located at $\sim$ 8 mbar for WASP-33b and $\sim$ 4 mbar for KELT-9b in the wavelength region studied here and assuming solar metallicity. Thus, for KELT-9b, \ion{Ca}{ii} is the dominant species in the region probed by transmission spectroscopy; for WASP-33b, \ion{Ca}{i} can be probed at lower altitudes but only within a small altitude range from 1 mbar to $\sim$ 8 mbar. 
We assumed isothermal temperature with $T_\mathrm{{eq}}$ values, but the actual temperature at the planetary terminators may deviate from $T_\mathrm{{eq}}$ depending on the 3D atmospheric circulation as well as other mechanisms such as temperature inversion. Therefore, the actual mixing ratio profiles could be different from the ones presented in Figure \ref{Mixing-ratio}.

We also searched for \ion{Ca}{i} in the two planets using the cross-correlation method. 
Since most of the \ion{Ca}{i} lines are in the HARPS-N wavelength region, we only cross-correlated the modeled \ion{Ca}{i} spectrum with the HARPS-N dataset. The simulated stellar RM and CLV effects were corrected before the cross-correlation process. We were not able to detect \ion{Ca}{i} signals in the two planets.
\cite{Hoeijmakers2019} searched for \ion{Ca}{i} in KELT-9b using two transit observations from HARPS-N and there was only tentative evidence of \ion{Ca}{i}. These observational results imply that the atmospheres of the two planets are extremely hot and \ion{Ca}{ii} is dominant.

%                                                Two column figure
%----------------------------------------------------------- S_vib
%-----------------------------------------------------------

%                                                Two column figure
%----------------------------------------------------------- S_vib
%-----------------------------------------------------------

\subsection{Model of \ion{Ca}{ii} transmission spectrum}
\begin{figure}
   \centering
   \includegraphics[width=0.49\textwidth]{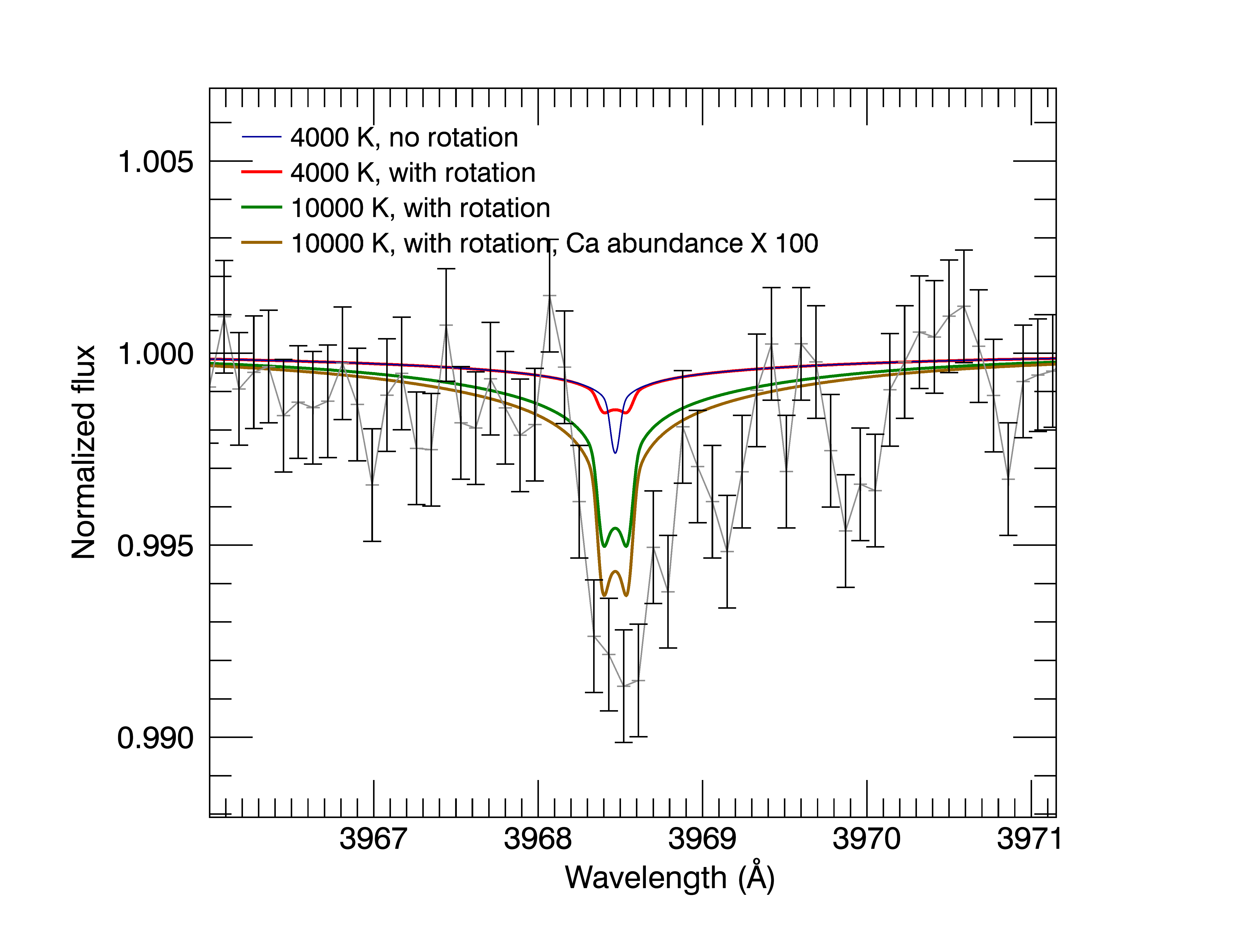}
      \caption{Different models for the \ion{Ca}{ii} H line absorption for KELT-9b. The black points are the observed transmission spectrum (binned every 7 points). The blue line is the model with a temperature of 4000 K. The red line is the model with rotational broadening included. The green line is the model with a temperature of 10000 K and the yellow line is with increased Ca abundance. The observed absorption is stronger than the model predictions. Such a strong absorption indicates a hydrodynamic outflow of the material. }
         \label{fig-K9-models}
   \end{figure}

The \ion{Ca}{ii} transmission spectra of both planets are significantly stronger than the model predictions that are calculated assuming equilibrium temperature and solar abundance (c.f. Figure \ref{fig-profiles}). The second last column of Table \ref{fit_result} lists the line depths from the models. Here we considered the rotational broadening by assuming tidal locking using the method of \cite{Brogi2016}. The observed line depths of WASP-33b are 8.6 and 10 times stronger than the depths from the model for the H\&K and IRT lines, respectively. For KELT-9b, the observed line depths are 4.8 and 3.8 times stronger for the H\&K and IRT lines, respectively. When increasing the isothermal temperature in the models, we obtained larger line depths because of the increased scale heights; the line depth also increases with an increasing calcium abundance. Figure \ref{fig-K9-models} compares different models for the H line in KELT-9b. Even with temperature as high as 10000 K and a Ca abundance 100 times the solar Ca abundance, the observed line depth is still stronger than the model.

\begin{figure}
   \centering
   \includegraphics[width=0.49\textwidth]{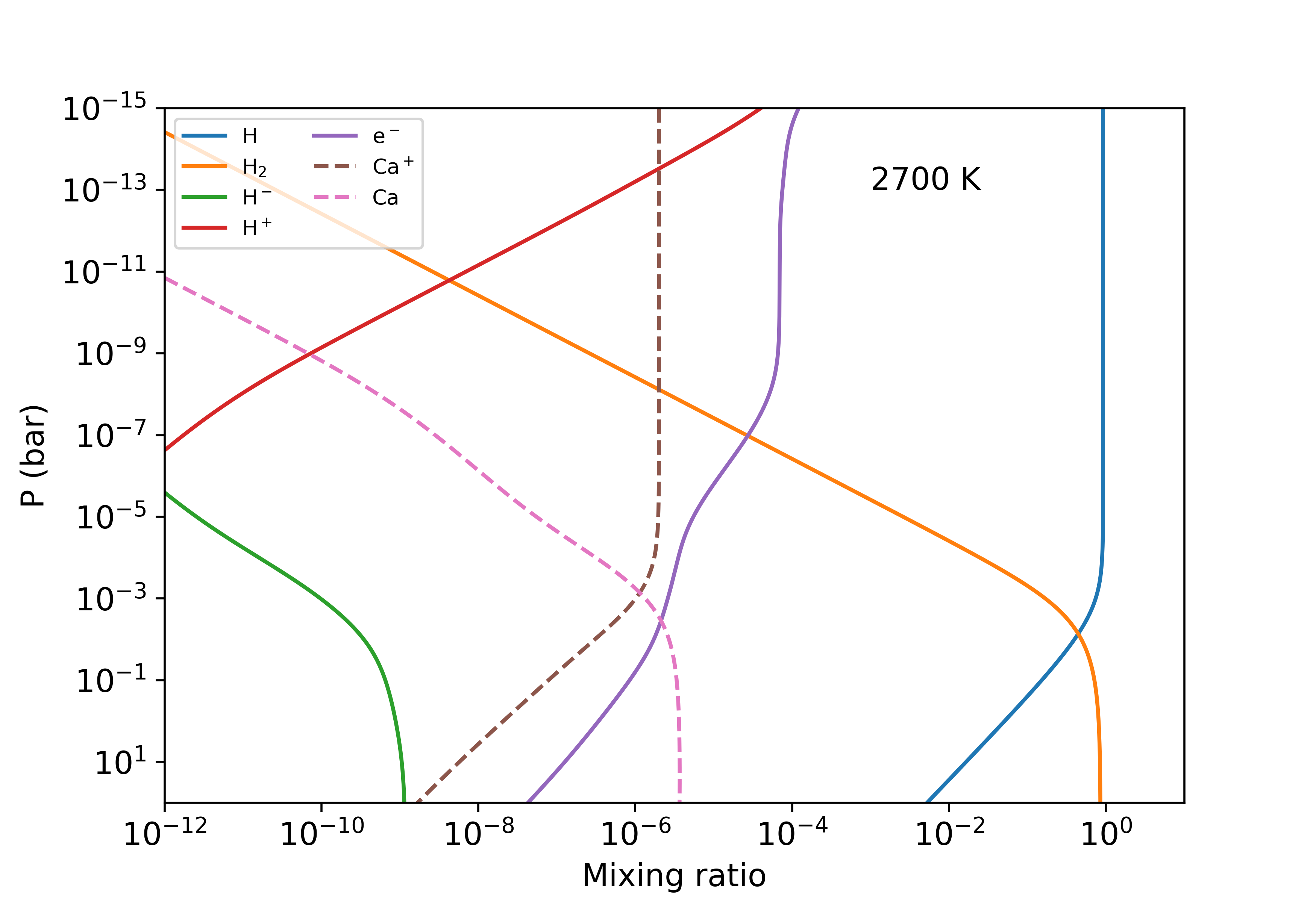}
   \includegraphics[width=0.49\textwidth]{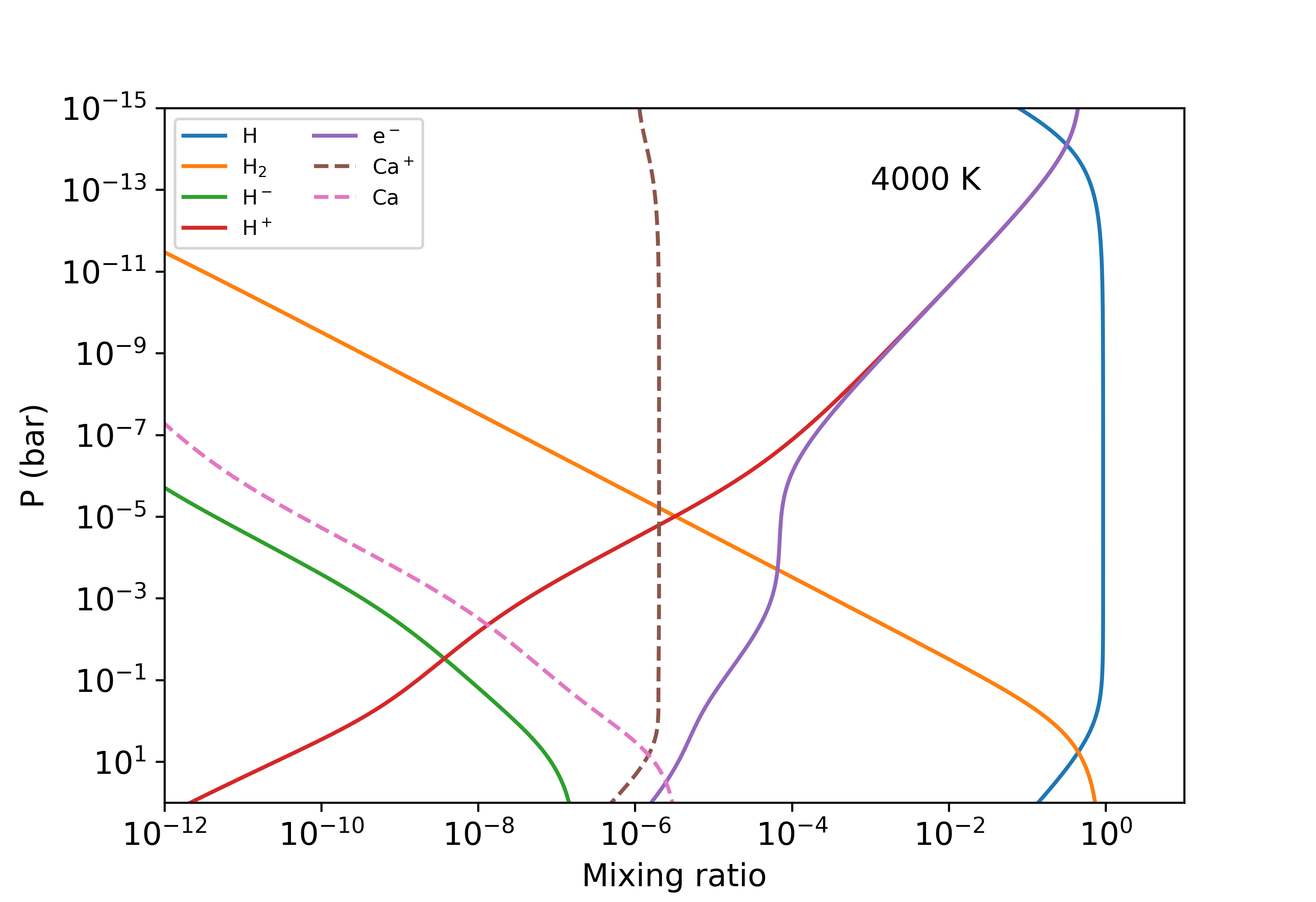}
      \caption{Mixing ratios of hydrogen species calculated assuming solar metallicity, equilibrium chemistry, and isothermal temperature. The upper panel is for WASP-33b (2700\,K) and the lower panel is for KELT-9b (4000\,K). In both planets, $\mathrm{H^-}$ exists at low altitudes and is the main opacity source of the spectral continuum. The calcium profiles are also plotted as dashed lines.
     }
         \label{H-mixing-ratio}
   \end{figure}

\begin{figure}
   \centering
   \includegraphics[width=0.49\textwidth]{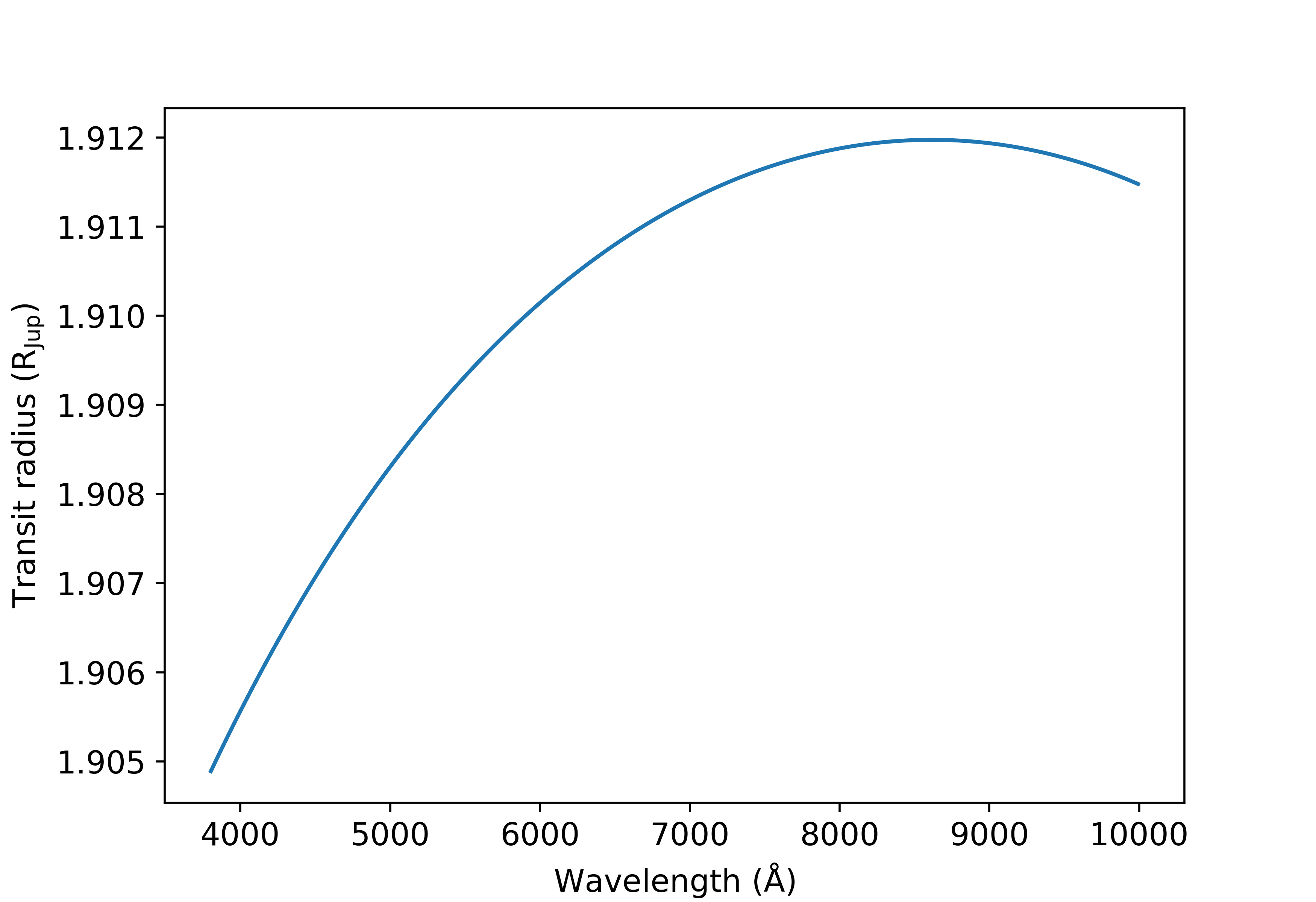}
      \caption{Modeled transmission spectrum of $\mathrm{H^-}$ for KELT-9b.}
         \label{H-spec}
   \end{figure}

When simulating the \ion{Ca}{ii} transmission spectrum, we set a cut-off level of 1 mbar to account for possible continuum opacities. The major continuum opacity for the two planets is $\mathrm{H^-}$ absorption. The $\mathrm{H^-}$ mixing ratio and the $\mathrm{H^-}$ transmission spectrum are presented in Figure \ref{H-mixing-ratio} and Figure \ref{H-spec}, respectively. The absorption of $\mathrm{H^-}$ peaks at $\sim$ 0.85 $\mathrm{\mu m}$. The average $\mathrm{H^-}$ continuum level in the optical wavelength range is around 4 mbar and 8 mbar for KELT-9b and WASP-33b, respectively. In order to evaluate the impact of choosing different continuum levels, we simulated \ion{Ca}{ii} transmission spectra using cut-off levels from 0.1 mbar to 10 mbar and found that the line depth increased by less than 20$\%$.
Therefore, we concluded that setting different continuum levels can not explain the strong \ion{Ca}{ii} lines observed.

Such a strong \ion{Ca}{ii} absorption in the two planets can be caused by the hydrodynamic escape that brings up calcium ions and, as a result, significantly enhances its density at high altitudes. Compared to hydrostatic models, the hydrodynamic outflow increases significantly the density of materials  at altitudes close to the Roche lobe \citep{Vidal-Madjar2004}.
\cite{Hoeijmakers2019} also found that the observed line depths are stronger than the modeled values and they attributed this discrepancy to the hydrodynamic escape that transports materials to high altitudes. 
Theoretical models predict that the atmospheres of UHJ are prone to strong atmospheric escape because they receive large amounts of stellar ultraviolet/extreme-ultraviolet radiation \citep{Fossati2018}. \cite{Yan2018} observed a strong $\mathrm{H\alpha}$ absorption in KELT-9b, which is evidence of substantial escape of the atmosphere.
Further \ion{Ca}{ii} line modeling work with hydrodynamic escape included will be able to constrain the temperature profile and mass loss rate of the planets \citep[e.~g.][]{Odert2019}.

%                                                Two column figure
%----------------------------------------------------------- S_vib
%-----------------------------------------------------------

%                                                Two column figure
%----------------------------------------------------------- S_vib
%-----------------------------------------------------------

%                                                Two column figure
%----------------------------------------------------------- S_vib
%-----------------------------------------------------------

\section{Conclusions}
We have detected singly ionized calcium in KELT-9b and WASP-33b -- the two hottest hot-Jupiters discovered so far. Together with the very recent \ion{Ca}{ii} detection in MASCARA-2b, these three UHJs are the only exoplanets with \ion{Ca}{ii} detected in their atmospheres. Our \ion{Ca}{ii} detections and lack of \ion{Ca}{i} detections demonstrate that calcium is probably mostly ionized into \ion{Ca}{ii} in the upper atmosphere of UHJs.

In addition to the detection using the cross-correlation method, we obtained the transmission spectra from the full set of the five \ion{Ca}{ii} lines (H\&K doublet and near-infrared triplet). The effective radii of the H\&K lines are close to the Roche lobes of the planets, indicating that the calcium ions are from the very upper atmospheres where mass loss is underway.
The obtained line depths are significantly stronger than predictions by hydrostatic models assuming an isothermal temperature of $T_\mathrm{{eq}}$. This is probably because the upper atmosphere is hotter than $T_\mathrm{{eq}}$ and hydrodynamic outflow brings up \ion{Ca}{ii} to the high altitudes. Further modeling work with hydrodynamic escape included is thought to be required to fit the line profile and retrieve the temperature structure.

Due to the high ionization rate of calcium in the upper atmospheres of UHJs and the strong opacities of the \ion{Ca}{ii} H\&K and near-infrared triplet lines, the \ion{Ca}{ii} transmission spectrum is especially suitable for probing the high altitude atmospheres and revealing the properties of this peculiar class of exoplanets. 
These lines have great potential for the study of planet-star interaction, such as atmospheric escape and the impact of stellar wind.

\begin{acknowledgements}
We are grateful to the anonymous referee for his/her report.
F.~Y. acknowledges the support of the DFG priority program SPP 1992 "Exploring the Diversity of Extrasolar Planets (RE 1664/16-1)".
CARMENES is an instrument for the Centro Astron\'omico Hispano-Alem\'an de
  Calar Alto (CAHA, Almer\'{\i}a, Spain). 
  CARMENES is funded by the German Max-Planck-Gesellschaft (MPG), 
  the Spanish Consejo Superior de Investigaciones Cient\'{\i}ficas (CSIC),
  the European Union through FEDER/ERF FICTS-2011-02 funds, 
  and the members of the CARMENES Consortium 
  (Max-Planck-Institut f\"ur Astronomie,
  Instituto de Astrof\'{\i}sica de Andaluc\'{\i}a,
  Landessternwarte K\"onigstuhl,
  Institut de Ci\`encies de l'Espai,
  Institut f\"ur Astrophysik G\"ottingen,
  Universidad Complutense de Madrid,
  Th\"uringer Landessternwarte Tautenburg,
  Instituto de Astrof\'{\i}sica de Canarias,
  Hamburger Sternwarte,
  Centro de Astrobiolog\'{\i}a and
  Centro Astron\'omico Hispano-Alem\'an), 
  with additional contributions by the Spanish Ministry of Economy, 
  the German Science Foundation through the Major Research Instrumentation 
    Programme and DFG Research Unit FOR2544 ``Blue Planets around Red Stars'', 
  the Klaus Tschira Stiftung, 
  the states of Baden-W\"urttemberg and Niedersachsen, 
  and by the Junta de Andaluc\'{\i}a.
  Based on data from the CARMENES data archive at CAB (INTA-CSIC).
  This work is based on observations made with the Italian Telescopio Nazionale Galileo (TNG) operated on the island of La Palma by the Fundaci\'on Galileo Galilei of the INAF (Istituto Nazionale di Astrofisica) at the Spanish Observatorio del Roque de los Muchachos of the Instituto de Astrofisica de Canarias. 
  P.~M and I.~S. acknowledge support from the European 82 Research Council under the European Union's Horizon 2020 research and innovation program under grant agreement No. 694513.
\end{acknowledgements}

%\clearpage %% delete if do not want to clear page before reference list

% for the bibliography, at the end
\bibliographystyle{aa} % style aa.bst

\bibliography{CaII-refer}

\end{document}